\documentstyle{ioplppt}
\begin{document}
\jl{6}
\begin{flushright}
University of Parma Preprint UPRF-95-438 \\[2mm]
December 1995
\end{flushright}

\title{
Gauging kinematical and internal symmetry groups 
for extended systems: the Galilean one-time and 
two-times harmonic oscillators\protect\ftnote{7}{This 
work has been supported partly
by I.N.F.N., Italy (iniziativa specifica FI-2) and partly
by the Network {\it Constrained Dynamical Systems}
of the E.U. Programme ``Human Capital and Mobility''.}
}[Gauging kinematical and internal dynamical symmetries.]

\author{R.~De Pietri\dag, L.~Lusanna\ddag\ 
and M. Pauri\dag\dag}
\address{\dag Dipartimento di Fisica - Sezione di Fisica Teorica,
Universit\`a di Parma, 43100 Parma, Italy 
and  I.N.F.N., Sezione di Milano, Gruppo Collegato di Parma.}
\address{\ddag I.N.F.N., Sezione di Firenze,
 Largo E. Fermi 2, Arcetri - 50127 Firenze - Italy.}
\address{\dag\dag Center for Philosophy of Science, University of Pittsburgh,
Pittsburgh PA, USA.\protect\ftnote{8}{On leave from 
Dipartimento di Fisica - Sezione di Fisica Teorica,
Universit\`a di Parma, 43100 Parma, Italy 
and  I.N.F.N., Sezione di Milano, Gruppo Collegato di Parma.}
}

\begin{abstract} 
The possible external couplings of an extended 
non-relativistic classical system are
characterized by {\it gauging} its maximal 
dynamical symmetry group 
{\it at} the {\it center-of-mass}. The Galilean one-time 
and two-times harmonic oscillators are exploited
as models.
The following remarkable results are then obtained: {\bf 1)} a 
peculiar form of interaction of the system {\it as a whole}
with the external {\it gauge} fields; 
{\bf 2)} a modification of the dynamical part
of the symmetry transformations, which is
needed to take into account the alteration of the dynamics
itself, induced by the {\it gauge} fields. In particular,
the Yang-Mills fields associated to the internal rotations
have the effect of modifying the
{\it time derivative} of the internal variables
in a scheme of minimal coupling (introduction of an {\it internal 
covariant derivative});
{\bf 3)} given their dynamical effect,  
the Yang-Mills fields associated to the 
internal rotations apparently define
a sort of {\it Galilean spin connection},
while the Yang-Mills fields associated to the 
quadrupole momentum and to the internal energy have the 
effect of introducing a sort of {\it dynamically induced
internal metric} in the relative space.
\end{abstract}
\maketitle

\def\nome#1{{ \label{#1} }}
\def\writenote#1{{ }}
\def\EqRef#1{{(\ref{#1})}}

\def\cita#1{{$^{\mbox{\scriptsize \cite{#1}}}$}}
\def\der#1#2{ {d #1 \over d #2} }
\def\partder#1#2{ {\partial #1 \over \partial #2} }
\def\funcder#1#2{ {\delta #1 \over \delta #2} }
\def\primato#1{{{#1}^\prime}}
\def\gTRE{\mbox{${^3}\! g$}}

\def\Be{\begin{equation}}
\def\Ba{\begin{array}}
\def\Ee{\end{equation}}
\def\Ea{\end{array}}
\def\Bea{\begin{eqnarray}}
\def\Eea{\end{eqnarray}}

\def\Ds{\displaystyle }
\def\Ts{\textstyle }
\def\Ss{\scriptstyle }
\def\SSs{\scriptscriptstyle }
\def\Sss{\scriptscriptstyle }

\def\OldRefP#1{{}}                              


\def\SqrtMuK{\sqrt{\frac{\mu}{k}}}

\def\smallhalf{{\scriptstyle{1\over2}}}
\def\half{{1\over2}}

\def\EL{\mbox{$\cal E\! L$}}
\def\deq{{\buildrel \rm \circ \over =}}
\def\PBeq{{\buildrel \rm ! \over = }}
\def\DEFeq{{\buildrel {\rm def} \over = }}

\def\BME{{\mbox{\boldmath $E$}}}
\def\BMH{{\mbox{\boldmath $H$}}}

\def\BMx{{\mbox{\boldmath $x$}}}
\def\BMp{{\mbox{\boldmath $p$}}}
\def\BMP{{\mbox{\boldmath $P$}}}
\def\BMpi{{\mbox{\boldmath $\pi$}}}
\def\BMr{{\mbox{\boldmath $r$}}}
\def\BMrho{{\mbox{\boldmath $\rho$}}}

\def\BMx{{\mbox{\boldmath $x$}}}
\def\BMy{{\mbox{\boldmath $y$}}}
\def\BMz{{\mbox{\boldmath $z$}}}

\def\Ac{{\cal A}}
\def\Bc{{\cal B}}
\def\Cc{{\cal C}}
\def\Dc{{\cal D}}
\def\Ec{{\cal E}}
\def\Fc{{\cal F}}
\def\Gc{{\cal G}}
\def\Kc{{\cal K}}
\def\Hc{{\cal H}}
\def\Ic{{\cal I}}
\def\Jc{{\cal J}}
\def\Lc{{\cal L}}
\def\Mc{{\cal M}}
\def\Pc{{\cal P}}
\def\Rc{{\cal R}}
\def\Sc{{\cal S}}
\def\Qc{{\cal Q}}
\def\Vc{{\cal V}}

\def\eps{{\varepsilon}}


\hyphenation{Di-par-ti-men-to}
\hyphenation{na-me-ly}
\hyphenation{al-go-ri-thm}
\hyphenation{pre-ci-sion}



\eqnobysec
\section{Introduction}

In this work we exploit a {\it gauge} technique 
introduced in previous papers \cite{DePietri95} to
characterize the possible external couplings of 
{\it extended  non-relativistic} dynamical
systems. The non-relativistic isotropic harmonic oscillator 
with {\it center-of-mass} is used as a model both in the standard
Newtonian description and in the presymplectic description
with two {\it first-class} constraints.

As well known, no {\it natural} prescription is 
available for carrying out
the coupling to external fields of an {\it extended} system
with certain conserved charges [mass, internal energy, intrinsic angular
momentum ({\it invariants} of the {\it extended Galilei} group%
\footnote{Note that, on the one hand, all the massive systems 
are realizations of the {\it extended} Galilei group, 
and, on the other, as shown in our previous papers 
\cite{DePietri95},  recourse to the {\it extended} group is a necessary
condition to implement the Utiyama procedure.}
), quadrupole momentum for the oscillator, Runge-Lenz vector for
the Kepler system, and so on).
Intuitively, one would expect that e.g. a gravitational field 
would probe the {\it individual} constituent masses (as it happens for
a dust of free particles); it is not clear, however, what to do
in case the constituents of the system are not free (cluster
of interacting particles). 

Since the only constructive and
self-consistent method to create a coupling to external fields,
in the case in which the conserved charges are connected to a
Lie algebra ($1^{st}$ Noether theorem), is the Utiyama procedure,
in this work we make the {\it ansatz} that the 
essential structural elements of the {\it spatial extension} of a dynamical
system are represented and summarized by its maximal
dynamical symmetry group viz. by the algebraic structure
of the whole set of the constants of the motion of the system.
This seems to be a natural development and enrichment of
the tratidional ``atomistic'' way of describing extended systems
in terms of constituent parts interacting among themselves.

The maximal dynamical symmetry group of the
isotropic harmonic oscillator with {\it center-of-mass}
is the semi-direct product of the {\it centrally extended} 
Galilei group and the $SU(3)\otimes U(1)$ group.
Actually, the explicit canonical realization\footnote{
see Section 2 and the Appendices.} 
of this maximal group is the {\it direct product} of a {\it singular}
realization of the $SU(3)\otimes U(1)$
group and the {\it singular scalar} realization of the
{\it centrally extended} Galilei group, 
corresponding to the {\it center-of-mass} of the
system \cite{Pauri}.

Thus, given the peculiar algebraic structure 
of the realization of the maximal dynamical symmetry,
we are led to implement our {\it ansatz} by 
{\it localizing} the {\it gauging} of the symmetry
at the {\it center-of-mass} of 
the system. We show thereby  that the
{\it gauge} procedure is meaningful also for {\it dynamical} symmetries 
besides the usual {\it kinematical} ones.

Let us remark that our {\it ansatz} seems suited for
clusters of interacting particles both in the case of confining
forces and of long-range interactions. A priori, 
this algebraic {\it ansatz} could be applied also to a dust of free
particles; this, however, would result in a non-standard
coupling of the external fields to the {\it center-of-mass}
and relative variables of the dust. It seems realistic to
retain the standard coupling to the dust, but use our {\it ansatz}
for the constituents of a dust of extended bound systems (for
instance a dust of harmonic oscillators).

In spite of the evident paradigmatic
and heuristic nature of our {\it ansatz}, the results
obtained here seem to be notably expressive.

The technical steps of the work are the following: {\bf 1})
the standard Utiyama procedure for fields is applied to
the possible trajectories of the {\it center-of-mass}
of the system as described by a canonical realization of the {\it extended} 
Galilei group. This determines the gravitational-inertial 
fields which can couple to the {\it center-of-mass} itself.
As shown elsewhere \cite{DePietri95}, the requirement
of invariance (properly {\it quasi-invariance}) of the Lagrangian
leads to the introduction of eleven {\it gauge}
compensating fields and their transformation properties.
{\bf 2})~The generalized Utiyama procedure is then applied
to the {\it internal} dynamical $SU(3)\otimes U(1)$ symmetry so that {\it gauge}
compensating fields have to be introduced in connection
to the {\it internal angular momentum} ({\it spin}), the {\it quadrupole
momentum} and the {\it internal energy}. As already indicated 
by the above structure, the crucial point is that, 
not only the internal transformations have a dynamical nature but,
also, that the Galilei
rotations subgroup affects the internal transformations so that 
a {\it semi-direct product} structure emerges.
 
The following remarkable results are then obtained: {\bf 1)} a 
peculiar form of interaction of the system {\it as a whole}
with the external {\it gauge} fields; 
{\bf 2)} a modification of the dynamical part
of the symmetry transformations, which is
needed to take into account the modification of the dynamics
itself induced by the {\it gauge} fields. In particular,
the Yang-Mills fields associated to the internal rotations
have the effect of modifying the
{\it time derivative} of the internal variables
in a scheme of minimal coupling (introduction of an {\it internal 
covariant derivative}).
Moreover, it is interesting to find that, given their 
dynamical effect,  the Yang-Mills fields associated to the 
internal rotations apparently define
a sort of {\it Galilean spin connection}. 
On the other hand, the Yang-Mills fields associated to the 
quadrupole momentum and to the internal energy have the 
effect of introducing a sort of {\it dynamically induced
internal metric} in the relative space,
together with the above mentioned modification of the {\it
interaction}.

These general features are reminiscent of some peculiar aspects
of the {\it relativistic string} dynamics, in particular of the 
so called {\it bootstrap} hypothesis, according to
which  the external fields that can couple to the 
system are all and the same which are already included 
in the field theory itself. Of course, the maximal 
symmetry group of the relativistic string is not a Lie
group but it is based instead on the infinite algebra of the
constants of motion studied by Pohlmeyer and Rehren \cite{PR}.
The specific study of this dynamical system will require 
a preliminary clarification of the case of the 
{\it relativistic} oscillator, in which a simple separation 
of the symmetry 
into {\it center-of-mass} and {\it internal} parts is 
no longer feasible.
Finally, the determination of the dynamical equations
for the Yang-Mills fields remains an open problem
and will be discussed elsewhere. 


\section{The one-time harmonic oscillator with center-of-mass}

The Lagrangian that describe the one-time harmonic
oscillator with {\it center-of-mass} is:
\Be
  \Lc = \sum_{a=1}^2 \frac{m_a}{2} \delta_{ij} \dot{x}_a^i\dot{x}_a^j
     -\frac{k}{2} \delta_{ij} (x_1^i - x_2^i )(x_1^j - x_2^j )
~~.
\nome{2.1}
\Ee
Introducing the {\it center-of-mass} and relative coordinates:
\Be
\left\{
\Ba{rcl}
   x^i &=& \Ds \frac{ m_1 x_1^i + m_2 x_2^i }{M} \\[3 mm]
   r^i &=& x_1^i - x_2^i ~~,
\Ea
\right.
\nome{2.2}
\Ee
the Lagrangian can be rewritten as:
\Be 
  \Lc = \frac{M}{2} \delta_{ij} \dot{x}^i \dot{x}^j
     +\frac{\mu}{2} \delta_{ij} \dot{r}^i\dot{r}^j
     -\frac{k}{2} \delta_{ij} r^i r^j
   ~\DEFeq~ \frac{M}{2} \delta_{ij} \dot{x}^i \dot{x}^j
          + \Lc_r (t)
~~,
\nome{2.3}
\Ee
where $\mu = \frac{m_1 m_2}{m_1 + m_2}$
is the reduced-mass and $M=m_1+m_2$ is the
total mass.

The canonical conjugate momenta are:
\Be
\left\{
\Ba{rcl}
 p_{1i} &\DEFeq& \Ds \partder{L}{\dot{x}_1^i}
           = m_1 \delta_{ij} \dot{x}_1^j
           = \Ds \frac{m_1}{M} P_i + \pi_i \\[3 mm]
 p_{2i} &\DEFeq& \Ds \partder{L}{\dot{x}_2^i}
           = m_2 \delta_{ij} \dot{x}_2^j
           = \Ds \frac{m_2}{M} P_i - \pi_i
\Ea
\right.
\Ee
where
\Be
\left\{
\Ba{rcl}
 P_i   &\DEFeq& \Ds \partder{L}{\dot{x}^i}
           = M \delta_{ij} \dot{x}^j
           =  p_{1i} + p_{2i} \\[3 mm]
 \pi_i &\DEFeq& \Ds \partder{L}{\dot{r}^i}
           = \mu \delta_{ij} \dot{r}^j
           = \Ds \frac{ m_2 p_{1i} - m_1 p_{2} }{M}  ~~.
\Ea
\right.
\nome{2.4}
\Ee
In phase space we have the following non-zero Poisson brackets:
\Be
\left\{
\Ba{rclcl}
  \{ x_1^i, p_{1j} \} &=& \{ x_2^i, p_{2j} \} &=& \delta_j^i  
\qquad a,b=1,2,3\\
  \{ x^i  , P_j \}    &=& \{ r^a  , \pi_b \}  &=& \delta_j^i  
\qquad i,j=1,2,3 ~, \\
\Ea
\right.
\nome{2.5}
\Ee
and the Hamiltonian results
\Be
\fl
\left\{
\Ba{rcl}
\bar{H}   &=& \Ds  \frac{1}{2m_1} \delta^{ij} p_{1i} p_{1j} 
       +\frac{1}{2m_2} \delta^{ij} p_{2i} p_{2j}
       +\frac{k}{2} \delta_{ij} (x_1^i -x_2^i) (x_1^j -x_2^j)
\\[2 mm]
          &=& \frac{1}{2M} \delta^{ij} P_i P_j + \bar{\Ec}  \; , 
\\[2 mm]
\bar{\Ec}  &\DEFeq&
             \Ds \frac{1}{2\mu} \delta^{ab} \pi_a \pi_b 
             +\frac{k}{2} \delta_{ab} r^a r^b   \; ,
\Ea
\right.
\nome{2.6}
\Ee
where $\bar{\Ec}$ is the internal energy
(from now on, we shall use the indexes $i,j,k,..$ for the 
{\it center-of-mass} components and $a,b,c,d,...$ for 
the {\it internal} components).
The notation $\bar{f}$ denotes a function on phase-space 
and the symbol $\PBeq$ the pull-back via the Legendre 
transformation. i.e. $\bar{f} (\BMx,\BMP;\BMr,\BMpi)
\PBeq f(\BMx,\dot{\BMx};\BMr,\dot{\BMr})$.
Before proceeding, let us fix some notations in connection with
more general classes of functions that will have to be considered.
Precisely: 
1) $f(\BMz , t)$, with $\BMz$ and $t$ independent variables; and
2) $f(\BMx(t),t)$.

The variations in case 1), when $t\rightarrow t^* = t + \delta t$,  
$\BMz\rightarrow \BMz^* = \BMz + \delta \BMz$, will be denoted
by
\Be
\Ba{rcl}
  \delta   f(\BMz,t) &=& \Ds f^*(\BMz^*,t^*) - f(\BMz,t) \\[1 mm]
  \delta_0 f(\BMz,t) &=& \Ds f^*(\BMz  ,t  ) - f(\BMz,t) = \\[3 mm]
                     &=& \Ds  \delta   f(\BMz,t)
                         - \partder{f(\BMz,t)}{z^k} \delta z^k    
                         - \partder{f(\BMz,t)}{t}   \delta t    ~~~.
\Ea
\Ee                            
On the other hand, in case 2), it is convenient to distinguish three
kinds of variations. Precisely, in correspondence to:
\Be
\fl
\left\{
\Ba{lcl}
t       &\rightarrow& t^* = t + \delta t  \\[1 mm]  
\BMx (t)&\rightarrow& \Ds \BMx^* (t^*) = \BMx (t) + \delta \BMx (t)
                     = \BMx (t) + \delta_0 \BMx (t) 
                      + \frac{d\BMx (t)}{dt} \delta t ~~, 
\Ea
\right.
\Ee
we shall define:
\Be
\fl
\Ba{rcl}
\delta   f(\BMx(t),t)&=& \Ds f^*(\BMx^*(t^*),t^*) - f(\BMx(t),t) 
\\[3 mm] 
\delta_0 f(\BMx(t),t)&=& \Ds f^*(\BMx(t)  ,t  ) - f(\BMx(t),t) = 
\\[3 mm]
            &=& \Ds \delta   f(\BMx(t),t)
              - \partder{f(\BMx(t),t)}{x^k}  \delta x^k(t)    
              - \partder{f(\BMx(t),t)}{t}    \delta t 
\\[3 mm]
\delta_{0[t]} f(\BMx(t),t) 
                   &=& \Ds f^*(\BMx^*(t)  ,t  ) - f(\BMx(t),t) = 
\\[3 mm]
                   &=& \Ds  \delta   f(\BMx(t),t)
                       - \left[
                          \partder{f(\BMx(t),t)}{x^k}  
                          \frac{dx^k(t)}{dt}    
                         +\partder{f(\BMx(t),t)}{t} 
                         \right]   \delta t ~~.\\
\Ea
\nome{2.7a}
\Ee                            

The generators of the Galilei group ${\Gc}$ for the 
particular realization of the harmonic oscillator can
be written  ($c_{ij}^{~~k} = \epsilon^{ijk}$ structure constants of
the SU(2) group):
\Be
\left\{
\Ba{rcl}
\bar{H}   &=& \Ds \frac{ \delta^{ij} P_i P_j }{ 2M } + \bar{\Ec} \\[2 mm]
P_i       &=& p_{1i} + p_{2i} \\[2 mm]
\bar{J}_i &=& c_{ij}^{~~k} x_1^j p_{1k} 
             +c_{ij}^{~~k} x_2^j p_{2k} \\[2 mm]
\bar{K}_i &=& m_1 \delta_{ij} x_1^j 
             +m_2 \delta_{ij} x_2^j  
             -t (p_{1i} + p_{2i} )    ~~,\\
\Ea
\right.
\nome{2.7}
\Ee
and they satisfy the {\it centrally extended}  
Galilei algebra with {\it central charge} $M$,
\Be
\left\{
\Ba{rcl}
  \{ \bar{H}  , \bar{K}_i \} &=& - P_i        \\
  \{ P_i      , \bar{K}_j \}     &=& - M \delta_{ij}  \\
  \{ P_i      , \bar{J}_j \}     &=& c_{ij}^{~~k} P_k  \\
  \{ \bar{J}_i, \bar{J}_j \}     &=& c_{ij}^{~~k} \bar{J}_k  \\
  \{ \bar{K}_i, \bar{J}_j \}     &=& c_{ij}^{~~k} \bar{K}_k  \\
  \{ \bar{K}_i, \bar{K}_j \}     &=& 0  ~~,
\Ea
\right.
\nome{2.8}
\Ee
The generator of a general  Galilei transformation is then
\Be
   \bar{G} =  \varepsilon \bar{H} + \varepsilon^i P_i
            + \omega^i \bar{J}_i + v^i \bar{K}_i ~~,
\nome{2.9}
\Ee
and the induced coordinate transformations are
\Be
\left\{
\Ba{rcl}
\bar{\delta}_0 x^i &=& \Ds \{ x^i , \bar{G} \}
                  = \epsilon ~\delta^{ij} \frac{P_j}{M} + \epsilon^i
                   +c_{jk}^{~~i} \omega^j x^k
                   - t v^i \\[3 mm]
\bar{\delta}_0 r^a &=& \Ds \{ r^a , \bar{G} \}
                  = \epsilon ~\delta^{ab} \frac{\pi_b}{\mu} 
                   +c_{bc}^{~~a} \omega^b r^c \\[3 mm]
\bar{\delta}_0 x_1^i &=& \Ds  \{ x_1^i , \bar{G} \}
                  = \epsilon \delta^{ij} \frac{p_{1j}}{m_1} + \epsilon^i
                   +c_{jk}^{~~i} \omega^j x_1^k
                   - t v^i    \\
\bar{\delta}_0 x_2^i &=& \Ds  \{ x_2^i , \bar{G} \}
                  = \epsilon ~\delta^{ij} \frac{p_{2j}}{m_2} + \epsilon^i
                   +c_{jk}^{~~i} \omega^j x_2^k
                   - t v^i   ~~.
\Ea
\right.
\nome{2.10}
\Ee
$\bar{G}$ gives rise to the following Noether transformations
\Be
\left\{
\Ba{rcl}
\delta_0 x^i   &=& \epsilon ~\dot{x}^i + \epsilon^i
                    +c_{jk}^{~~i} \omega^j x^k
                    - t v^i \\[1 mm]
\delta_0 r^a   &=& \epsilon ~\dot{r}^a + c_{cd}^{~~a} \omega^b r^c \\[1 mm]
\delta_0 x_1^i &=& \epsilon ~\dot{x}_1^i + \epsilon^i
                   +c_{jk}^{~~i} \omega^j x_1^k
                   - t v^i   \\
\delta_0 x_2^i &=& \epsilon ~\dot{x}_2^i + \epsilon^i
                   +c_{jk}^{~~i} \omega^j x_2^k
                   - t v^i  ~~,
\Ea
\right.
\nome{2.11}
\Ee
under which the Lagrangian is {\it quasi-invariant}. In fact:
\Be
\Ba{rcl}
\delta_0 \Lc &=& \Ds \partder{\Lc}{x^i} \delta x^i 
              +      \partder{\Lc}{r^a} \delta r^a
              +      \partder{\Lc}{\dot{x}^i} \frac{d}{dt} \delta x^i 
              +      \partder{\Lc}{\dot{r}^a} \frac{d}{dt} \delta r^a   
\\
         &=& \Ds \frac{d}{dt}\left[ 
                     \epsilon \Lc  - M \delta_{ij} v^i x^j
                         \right] ~~.
\Ea
\nome{2.12}
\Ee
Therefore, the Galilei generators are constants of the motion and
the Galilei group is a kinematical  symmetry group of the system.

As well known, in the case of a harmonic oscillator,
we have an additional dynamical symmetry
group, viz. $SU(3)\otimes U(1)$, whose real form 
realization has the following generators
\Be
\fl
\left\{
\Ba{rcl}
  \bar{\Ec}^\prime  &=& \SqrtMuK \bar{\Ec}
       =  \frac{1}{2} \left(
                 \frac{1}{\sqrt{\mu k}} \delta^{ab} \pi_a \pi_b 
               + \sqrt{\mu k} \delta_{ab} r^a r^b
                \right) 
\\[2 mm]
      &\PBeq& \frac{1}{2} \left(
                \frac{\mu^2}{\sqrt{\mu k}} \delta_{ab} \dot{r}^a \dot{r}^b 
                       + \sqrt{\mu k} \delta_{ab} r^a r^b
                        \right)
\\[2 mm]
   \bar{S}_{a}       &=& (\vec{r} \wedge \vec{\pi} )_a
                      =   c_{ab}^{~~c} r^b \pi_c
                    \PBeq \mu c_{ab}^{~~c} r^b 
                          \delta_{cc_1} \dot{r}^{c_1} \\[2 mm]
  \bar{\Qc}_{ab}    &=& \Ds \frac{1}{2 \sqrt{\mu k}} \pi_a \pi_b
                      +\frac{\sqrt{\mu k}}{2} 
                        \delta_{ac} \delta_{bd} r^c r^d
                      -\delta_{ab}\frac{\bar{\Ec}^\prime}{3} 
\\[3 mm]
                    &\PBeq&\Ds  \delta_{ac} \delta_{bd} \left[
                        \frac{\mu^2}{2 \sqrt{\mu k}} 
                        \dot{r}^c \dot{r}^d
                       +\frac{\sqrt{\mu k}}{2}   r^c r^d   \right]
                       -\delta_{ab}\frac{\Ec^\prime}{3} 
\; .
\Ea
\right.
\nome{2.13}
\Ee
Here  $\bar{\Ec}$  is the internal energy, 
$\bar{S}_a$ the internal angular  momentum (``spin'') and
$\bar{\Qc}_{ab}$ the quadrupole moment.
$\bar{\Ec}$ and $\bar{\Qc}_{ab}$ may be described in a more compact
form by the symmetric tensor
\Be
\Ba{rcl}
  \bar{N}_{ab} &=& \Ds \frac{1}{2 \sqrt{\mu k}} \pi_a \pi_b
                  +\frac{\sqrt{\mu k}}{2}  
                   \delta_{ac} \delta_{bd} r^c r^d    \\[3 mm]
               &=& \Ds \bar{\Qc}_{ab} 
                   + \delta_{ab} \frac{\bar{\Ec}^\prime }{3} ~~.
\Ea
\nome{2.14}
\Ee
A real form of the $SU(3)\otimes U(1)$ algebra is
\Be
\left\{
\Ba{rcl}
   \{ \bar{S}_a   , \bar{S}_b \} &=&  c_{ab}^{~~c} \bar{S}_c    \\
   \{ \bar{S}_a   , \bar{N}_{cd} \} &=& c_{ac}^{~~b} \bar{N}_{bd}
                                       +c_{ad}^{~~b} \bar{N}_{cb} \\
   \{ \bar{N}_{ab}, \bar{N}_{cd} \} &=& \frac{1}{4} [
           \delta_{bd} c_{ac}^{~~e}+ \delta_{bc} c_{ad}^{~~e}
         + \delta_{ad} c_{bc}^{~~e}+ \delta_{ac} c_{bd}^{~~e}] \bar{S}_e ~~.
\Ea
\right.
\nome{2.15}
\Ee
Some other properties of the canonical $SU(3)\otimes U(1)$ 
realizations are reported in Appendix A.

The general $SU(3)\otimes U(1)$ transformations generated by
$\tilde{G} ~\DEFeq~ \theta^a \bar{S}_a 
+ \xi^{\prime ab} \bar{\Qc}_{ab} + {\xi}^{\prime} \Ec'$
or equivalently 
\Be
  \tilde{G} ~\DEFeq~ \theta^a \bar{S}_a 
    + \xi^{ab} \bar{N}_{ab} ~~,
\Ee
induce the following Noether transformations
\Be
\left\{
\Ba{rclcl}
  \bar\delta_0 x^i &\PBeq& \delta_0 x^i &=& 0 \\[2 mm]
  \bar\delta_0 r^a &\PBeq& \delta_0 r^a 
                   &=& \Ds c_{bc}^{~~a} \theta^b r^c
          + \SqrtMuK \xi^{ab} \delta_{bc} \dot{r}^c ~~,
\Ea
\right.
\nome{2.16}
\Ee
under which the Lagrangian is {\it quasi-invariant}:
\Be
\fl
\Ba{rcl}
 \delta_0 \Lc &=&\Ds \partder{\Lc}{r^a} \delta_0 r^a
           +\partder{\Lc}{r^a} \frac{d}{dt} \delta_0 r^a
           = \frac{d}{dt} \left[ \sqrt{\frac{\mu}{k}} \xi^{ab}
                              \delta_{ac} \delta_{bd}
                              \left( \frac{\mu}{2} \dot{r}^c \dot{r}^d
                                    -\frac{k}{2} r^c r^d
                              \right)\right]
\\[2 mm]
          &\equiv& \frac{d}{dt} \left[ \sqrt{\frac{\mu}{k}} \xi^{ab}
                              F_{ab}( \dot{\BMr};\BMr )
                              \right] ~~.
\Ea
\nome{2.17}
\Ee
Therefore the $SU(3)\otimes U(1)$ generators are constants of the motion too.

Under the transformations \EqRef{2.16}, the generators 
$\bar{S}_a,\bar{N}_{ab}$
behave in the following way:
\Be
\fl
\left\{
\Ba{rcl}
\bar{\delta}_0 \bar{S}_a
    &=& \{ \bar{S}_a, \bar{\tilde{G}} \}
    = \Ds c_{ab}^{~~c} \theta^b  \bar{S}_c
       +\xi^{cd} c_{ac}^{~~b} \bar{N}_{bd}  
       +\xi^{cd} c_{ad}^{~~b} \bar{N}_{cb}  
\\[2 mm]
    &\PBeq& \Ds \delta_0 S_a - \sqrt{\frac{\mu}{k}} 
                 c_{ab}^{~~c} \xi^{bd} \delta_{ce}\delta_{df} r^e
       \left[ \frac{\delta\Lc}{\delta r^f}
             -\frac{d}{dt} \frac{\delta\Lc}{\delta\dot{r}^f} \right]
      \deq \delta_0 S_a \\[2 mm]
\bar{\delta}_0 \bar{N}_{ab} 
    &=& \{ \bar{N}_{ab}, \tilde{\bar{G}} \}
     =  \Ds c_{ac}^{~~d} \theta^c  \bar{N}_{db}
       +    c_{bc}^{~~d} \theta^c  \bar{N}_{ad}
       +\frac{1}{2} ( c_{ac}^{~~e} \delta_{bd} \xi^{cd}
                     +c_{bc}^{~~e} \delta_{ad} \xi^{cd})\bar{S}_{e}  \\[2 mm]
    &\PBeq& \Ds \delta_0 N_{ab}
     +\frac{\mu}{2k} \delta_{ac} \delta_{bd} \delta_{ef}
       ( \dot{r}^c \xi^{de}+\dot{r}^d \xi^{ce} )
       \left[ \frac{\delta\Lc}{\delta r^f}
             -\frac{d}{dt} \frac{\delta\Lc}{\delta\dot{r}^f} \right]
      \deq \delta_0 N_{ab} ~~.
\Ea
\right.
\nome{2.18}
\Ee

By ``putting together'' the {\it central extension} of the Galilei 
algebra and the $SU(3)\otimes U(1)$
algebra, we obtain the Lie algebra $\Gc_\Hc$ of the 
{\it maximal symmetry group}
$\Hc$ whose  Poisson brackets  are given by eqs \EqRef{2.8},\EqRef{2.15} and
\Be
\left\{
\Ba{rcl}
  \{ \bar{S}_a    , \bar{J}_i \} &=& \epsilon_{aic} \bar{S}_c \\
  \{ \bar{N}_{ab} , \bar{J}_i \} &=& \epsilon_{aic} \bar{N}_{cb}
                                    +\epsilon_{bic} \bar{N}_{ac}
~~.
\Ea
\right.
\nome{2.19}
\Ee
We get in this way a {\it semi-direct} product structure for the maximal
symmetry group:  $\Hc = (\Gc_M \wedge SU(3)) \otimes U(1)$.

Let us remark that, for a generic spin-independent central
interaction, the ``internal'' dynamical symmetry group is 
reduced from $SU(3)\otimes U(1)$ to
$SU(2) \otimes U(1)$, spanned the internal energy 
$\bar{\Ec} = \Ec / \mu$ and the
internal angular momentum $S_a$. Only in special cases
the symmetry group is larger than $SU(2)\otimes U(1)$.
As is well known, for instance, in the case of Kepler
problem, $SU(2)\otimes U(1)$ is enlarged to 
$SO(4) \otimes U(1)$ for the closed orbits 
(attractive force, $\Ec < 0$), to $E(3) \otimes U(1)$ 
($E(3)$ Euclidean group) for parabolic motions (with $\Ec=0$), 
and to $SO(3,1) \otimes U(1)$ for hyperbolic motion ($\Ec > 0$). 
In the Kepler case, the
generators are the internal energy $\bar{\Ec}$, the spin $S_a$ and
either $\vec{K} = \vec{R} /\sqrt{2|\Ec|} \;\; (\Ec \neq 0)$ 
or $\vec{K^\prime} = \vec{R}$ $(\Ec =0)$, where 
$\vec{R}= \vec{\pi}\wedge\vec{S} / \mu - G \mu M \vec{r}/ r$ 
is the Runge-Lenz vector constant of the motion
($\mu$ reduced mass, $M$ total mass, $G$ graviational constant).
Note that for two free particles, the ``internal''
dynamical symmetry group becomes $U(1)\otimes E(3)$
and the Euclidean transformations are generated 
by the relative momentum $\vec{\pi}$.


\section{Gauging the $SU(3)\otimes U(1)$ dynamical 
algebra in the one-time theory}

We have now to implement the idea of describing the
{\it extension} of the dynamical system by means 
of both a {\it point-like} structure
identified with the {\it center-of-mass} and an {\it internal}
structure summarized by its {\it dynamical
symmetry}.

We shall proceed as follows. First of all, we
observe that Newton's mechanics of a point particle may be
described using a trivial configuration bundle defined by
the {\it absolute-time} axis, as {\it base} manifold, and the
configuration space $R^3$, as {\it fiber}. 
The corresponding geometrical structure for a system of two particles
having a natural center of mass ($\BMx$) and relative
variables ($\BMr$), can still be characterized by a
trivial fiber bundle defined by the {\it absolute-time}
axis as {\it base} manifold. Yet, the {\it fiber} 
can be expediently viewed as a
fiber bundle ($Q$) by itself, having its {\it base} defined by 
a vector space $R^3$ ($Q_x$), associated to $\BMx$, and a {\it fiber}
associated to the relative or {\it internal} variable $\BMr$  ($Q_r$).
Introducing the dynamics requires that the configuration
space $Q$ with coordinates $\{ \BMx , \BMr \}$ be replaced by
its tangent bundle $TQ$ with coordinates $\{ \BMx ,
\BMr ; \dot{\BMx} ,\dot{\BMr} \}$. In this way  we get 
a further  trivial bundle over the {\it absolute-time} axis. 
Therefore, since the dynamical algebra acts only on the $\Lc_r(t)$
part of the Lagrangian (see eq (\ref{2.3})) with $\Lc_r
: TQ_r \mapsto R$, in carrying out the {\it gauging} of 
the internal dynamical group one is naturally led to replace the
arbitrary constants $\theta^a$, $\xi^{ab}$ of the Noether
transformations (\ref{2.16}) with arbitrary functions
$\theta^a(\BMx,t)$, $\xi^{ab}(\BMx,t)$ of the 
{\it center-of-mass} coordinates and the {\it absolute-time}.

In these conditions, one should expect that, in order the 
Lagrangian to remain {\it quasi-invariant} under
the localized $SU(3)\otimes U(1)$ transformations
as in eq. (2.22), compensating $SU(3)\otimes U(1)$
Yang-Mills fields, depending on the {\it center-of-mass}
coordinates $\BMx$, have to be introduced.

We will work out this program by a step-wise procedure.
First of all, let us observe that, under the
{\it internal} transformations
\Be
\left\{
\Ba{rcl}
  \delta^g_0 x^i &=& 0 \\
  \delta^g_0 r^a &=& \Ds \theta^b(\BMx) c^{~~a}_{bc} r^c
                +\sqrt{\frac{\mu}{k}} 
                 \xi^{ab} (\BMx) \delta_{bc} \dot{r}^c \; ,
\Ea
\right.
\nome{3.1}
\Ee
it follows
\Be
\Ba{rcl}
\delta^g_0 {\cal L} &=& \Ds \frac{d}{dt}
          \left[\sqrt{\frac{\mu}{k}} \xi^{ab} (x) 
                \delta_{ac}\delta_{bd}
                \left({ \frac{\mu}{2} \dot{r}^c\dot{r}^d 
                       -\frac{k}{2} r^c r^d }\right)
          \right]  \\[2 mm]
   & & + \Ds \dot{x_k}\partder{\theta^a}{x^k} ~S_a(\BMr,\dot{\BMr})
       + \dot{x_k}\partder{\xi^{cd}}{x^k} ~N_{cd}(\BMr,\dot{\BMr})
\; ,
\Ea
\nome{3.2}
\Ee
where the function $S_a(\BMr,\dot{\BMr})$,
$N_{ab}(\BMr,\dot{\BMr})$ are defined 
in eqs. \EqRef{2.13} and \EqRef{2.14}. 
The difference between the present 
case and the ``colored'' pseudo-classical
particle case is that in the former the Lie algebra 
generators are realized dynamically in terms
of relative variables  of the bi-local system, while in 
the latter they are realized in term of Grassmann
variables describing the {\it color} charge degrees 
of freedom of the point like-system in 
a ``quasi-classical'' approximation of the
standard first-quantized case \cite{GpP}.

 Then, since eqs. \EqRef{3.1} imply:
\Be
\Ba{rcl}
\delta^g_0 \dot{r}^a  
  &=& \Ds  \theta^b (\BMx) c_{bc}^{~~a} \dot{r}^c
          + \sqrt{\frac{\mu}{k}} \xi^{ab}(\BMx) 
            \delta_{bc} \ddot{r}^c  
\\[3 mm]
  & & \Ds + \dot{x}^k \partder{\theta^b}{x^k} c_{bc}^{~~a} {r}^c
          + \sqrt{\frac{\mu}{k}} \dot{x}^k\partder{\xi^{ab}}{x^k}
            \delta_{bc} \dot{r}^c \, ,
\Ea
\nome{3.3}
\Ee
eqs.\EqRef{2.18} are now replaced by
\Be
\fl
\left\{
\Ba{rcl}
{\delta}_0^g {S}_a  &=& \Ds c_{ab}^{~~c} \theta^b {S}_c
       +\xi^{cd} c_{ac}^{~~b} {N}_{bd}  
       +\xi^{cd} c_{ad}^{~~b} {N}_{cb}  
\\[2 mm]
    & &\Ds  + \sqrt{\frac{\mu}{k}} 
                 c_{ab}^{~~c} \xi^{bd} \delta_{ce}\delta_{df} r^e  
       \left[ \frac{\delta\Lc}{\delta r^f}
             -\frac{d}{dt} \frac{\delta\Lc}{\delta\dot{r}^f} \right]
\\[2 mm]
    & &\Ds  + {\mu} \dot{x}^k \theta^b_{,k}
              \delta_{cd} c_{ae}^{~~c} c_{bf}^{~~d}
              r^e r^f
            + \frac{\mu^2}{\sqrt{\mu k}} \dot{x}^k \xi^{cd}_{,k}
              \delta_{ce} \delta_{df} c_{ab}^{~~e} 
              r^b \dot{r}^f
\\[2 mm]
{\delta}_0^g {N}_{ab}  
    &=& \Ds c_{ac}^{~~d} \theta^c  \bar{N}_{db}
       +    c_{bc}^{~~d} \theta^c  \bar{N}_{ad}
       +\frac{1}{2} ( c_{ac}^{~~e} \delta_{bd} \xi^{cd}
                     +c_{bc}^{~~e} \delta_{ad} \xi^{cd})\bar{S}_{e}  \\[2 mm]
    & &\Ds -\frac{\mu}{2k} \delta_{ac} \delta_{bd} \delta_{ef}
       ( \dot{r}^c \xi^{de}+\dot{r}^d \xi^{ce} )    
       \left[ \frac{\delta\Lc}{\delta r^f}
             -\frac{d}{dt} \frac{\delta\Lc}{\delta\dot{r}^f} \right]
\\[2 mm]
    & &\Ds  + \frac{\mu}{\sqrt{k}} \dot{x}^k \theta^c_{,k}
              \delta_{af} \delta_{bg} c_{ce}^{~~f} 
              r^e \dot{r}^g
            + \frac{\mu^2}{{k}} \dot{x}^k \xi^{cd}_{,k}
              \delta_{ac} \delta_{be} \delta_{df}
              \dot{r}^e \dot{r}^f  ~~,
\Ea
\right.
\nome{3.4}
\Ee
where essentially new terms appear which  contain the space derivatives of 
the group parameters. Let us remark that, were it not for
the presence of these derivative terms, the introduction of
Yang-Mills {\it compensating} fields in the simple form:
\Be
  {\Lc} \rightarrow {\Lc}^\prime \equiv
           {\Lc} + \dot{x}^k B_k^{(a)} (\BMx) S_a (\BMr,\dot{\BMr})
                 + \dot{x}^k B_{k}^{(ab)} (\BMx) N_{ab} (\BMr,\dot{\BMr})
\; ,
\nome{3.5}
\Ee
would be already effective to obtain the desired result. In fact,
provided that the standard transformation properties be assumed
for the Yang-Mills fields $B^{(a)}_k$ and $B^{(ab)}_k$,
\Be
\left\{
\fl
\Ba{rcl}
  \delta B_{k}^{(a)}(\BMx)
      &=& \Ds   c_{bc}^{~~a} \theta^b (\BMx) B_{k}^{(c)}(\BMx) 
        + \xi^{cd} (\BMx)
            [c_{ce}^{~~a} \delta_{df} B_{k}^{(ef)}(\BMx)
            +c_{de}^{~~a} \delta_{cf} B_{k}^{(ef)}(\BMx)  ]  
\\[2 mm]
             & & \Ds  - \partder{\theta^a (\BMx)}{x^k}  
\\[3 mm]
  \delta B_{k}^{(ab)}(\BMx)
           &=&\theta^c (\BMx) [ c_{cd}^{~~a} B_{k}^{(db)} (\BMx)
                               +c_{cd}^{~~b} B_{k}^{(ad)} (\BMx) ]  
\\[2 mm]
           & &\Ds   + \frac{1}{4} \xi^{cd}(\BMx)
                 [c_{ec}^{~~a} \delta_d^b
                 +c_{ec}^{~~b} \delta_d^a
                 +c_{ed}^{~~a} \delta_c^b
                 +c_{ed}^{~~b} \delta_c^a] B_{k}^{(e)}(\BMx) 
\\[2 mm]
           & &\Ds - \partder{\xi^{ab}(\BMx)}{x^k} ~~,
\Ea
\right.
\nome{3.7}
\Ee
we would have,
\Be
\fl
\Ba{rcl}
\delta_0^g \Lc^\prime &=& \Ds \frac{d}{dt} 
         \left[ \sqrt{\frac{\mu}{k}} \xi^{ab}
         \delta_{ac} \delta_{bd}
        \left( \frac{\mu}{2} \dot{r}^c \dot{r}^d
        +\frac{k}{2\mu} r^c r^d
         \right)\right]   \\[2 mm]
       & &+ \Ds \dot{x}^k \left[ 
              \sqrt{\frac{\mu}{k}} B_k^{(a)} 
          c_{ab}^{~~c} \xi^{bd} \delta_{ce}\delta_{df} r^e  
        - \frac{\mu}{2k} B_{k}^{(ab)} \delta_{ac} \delta_{bd} \delta_{ef}
       ( \dot{r}^c \xi^{de}+\dot{r}^d \xi^{ce} )    
                         \right] \cdot \\[2 mm]
       & & \Ds ~~~~~\cdot \left[ \frac{\delta\Lc}{\delta r^f}
             -\frac{d}{dt} \frac{\delta\Lc}{\delta\dot{r}^f} \right]
\Ea
\nome{3.6}
\Ee
i.e. a {\it quasi-invariance} modulo the equations of
motion  $\frac{\delta\Lc}{\delta r^f}
-\frac{d}{dt} \frac{\delta\Lc}{\delta\dot{r}^f} \deq 0$
for the quadrupole part.

Therefore, if the Yang-Mills transformations \EqRef{3.7} 
have to be preserved, extra terms have to be added to the
Lagrangian in order to  compensate the extra terms
containing $\partial\theta^a (\BMx)/\partial x^k$,
$\partial\xi^{ab} (\BMx)/\partial x^k$ in eqs.\EqRef{3.4} 
which arise from the dynamical character of the $SU(3)\otimes U(1)$ 
realization. 

In order to identify these extra terms,
let us confine, as a first step, to the subgroup SU(2)
only ($\xi^{ab}=0$). In this case, it is evident that 
the extra terms must have the effect of replacing the 
time derivatives of the relative 
variables ($\dot{r}^a$) with the covariant derivative
\Be
 D r^a = \dot{r}^a + \dot{x}^k B_{k}^{(b)} (\BMx) c_{bc}^{~~a} r^c
~~.
\nome{3.8}
\Ee
Then, the {\it quasi-invariant} Lagrangian would result
\Be
 \Lc_{SU(2)} = \frac{M}{2} \delta_{ij} \dot{x}^i \dot{x}^j 
            -\frac{k}{2\mu} \delta_{ab} {r}^a {r}^b
            +\frac{\mu}{2} \delta_{ab} D r^a ~ D r^b  \; .
\nome{3.9}
\Ee
Since the momenta are
\Be
\left\{
\Ba{rcl}
   p_i   &=& M  \delta_{ij} \dot{x}^j 
            +\mu B_i^{(b)}(\BMx) c_{bc}^{~~a} r^c \delta_{ad} Dr^d \\
   \pi_a &=& \mu  \delta_{ab}  Dr^b  ~~,
\Ea
\right.
\nome{3.10}
\Ee
we would get a {\it minimal coupling} in the Hamiltonian:
\Be
\fl
\Ba{rcl}
\bar{H}_{SU(2)} &=& \Ds \frac{1}{2M}\delta^{ij}   
    \left[ P_i - B_{i}^{(a)} (\BMx) S_a(\BMr,\BMpi) \right]  
    \left[ P_j - B_{j}^{(a)} (\BMx) S_a(\BMr,\BMpi) \right] 
\\[2 mm] 
                & & \Ds +\frac{1}{2\mu} \delta^{ab} \pi_a \pi_b 
                   +\frac{k}{2\mu}  \delta_{ab} r^a r^b
\; .
\Ea
\nome{3.11}
\Ee
Note that $\bar{H}_{SU(2)}$ is indeed invariant
under the phase-space transformations generated by
$\theta^a (\BMx) \bar{S}_a$ if $\delta B_k^{(a)} (\BMx) $
is given by eqs.\EqRef{3.7} with $\xi^{ab} (\BMx) =0$.

  It is now straightforward to check, via eqs.\EqRef{3.7},
that the generalized minimally coupled Hamiltonian
\Be
\fl
\Ba{rcl}
\bar{H}_{SU(3)\otimes U(1)} &=& \Ds \frac{1}{2M} \delta^{ij} 
           \left[ P_i - B_{i}^{(a)}(\BMx)  S_a
                      - B_{i}^{(ab)}(\BMx) N_{ab} \right] \\[2 mm]
      & & ~~~~~~~~ \left[ P_i - B_{j}^{(a)}(\BMx)  S_a
                    - B_{j}^{(ab)}(\BMx) N_{ab} \right] \\[2 mm] 
               & &\Ds +\frac{1}{2\mu} \delta^{ab} \pi_a \pi_b 
                  +\frac{k}{2} \delta_{ab} r^a r^b 
\Ea
\nome{3.12}
\Ee
is invariant under the complete  $SU(3)\otimes U(1)$ 
transformations generated by
$\theta^a (\BMx) \bar{S}_a + \xi^{ab} (\BMx) \bar{N}_{ab} $.
Actually, we have 
\Be
\bar{\delta}_0 P_i =-\partder{\theta^a (\BMx)}{x^i} \bar{S}_a
                    -\partder{\xi^{ab} (\BMx)}{x^i} \bar{N}_{ab}
~~,\nome{3.13}
\Ee
so that  
$\bar{\delta}_0 [P_k - B^{(a)}_k (\BMx) S_a - B^{(ab)}_k (\BMx) N_{ab}] =0$.
Then, since 
$$ 
 \bar{\delta}_0 \left[\frac{1}{2\mu} \delta^{ab} \pi_a \pi_b 
+\frac{k}{2} \delta_{ab} r^a r^b \right] = 0 ~~,
$$
$\bar{H}_{SU(3)\otimes U(1)}$ is seen to be strictly invariant.

\vspace{1mm}
 To obtain the {\it quasi-invariant} Lagrangian 
$\Lc_{SU(3)\otimes U(1)}$, we have 
simply to perform the inverse Legendre transformation
defined via the first half of Hamilton's equations
\Be
\left\{
\Ba{rcl}
 \dot{x}^i  \deq \Ds \partder{\bar{H}}{P_i}
            &=& \Ds \frac{1}{M} \delta^{ij} \left[ 
                         P_j - B_{j}^{(a)}(\BMx) \bar{S}_a
                             - B_{j}^{(ab)}(\BMx) \bar{N}_{ab}
                          \right] \\[3 mm]
 \dot{r}^a  \deq \Ds \partder{\bar{H}}{\pi_a}
       &=& \Ds \delta^{ab} \frac{\pi_b}{\mu}
           - \partder{\bar{H}}{P_k}
             \left[ B_{k}^{(c)}(\BMx) \partder{\bar{S}_c   }{\pi_a}
                   +B_{k}^{(cd)} (\BMx)\partder{\bar{N}_{cd}}{\pi_a}
             \right]  \\[4 mm]
       &=& \Ds \delta^{ab} \frac{\pi_b}{\mu}
           -\dot{x}^k
           \left[ B_{k}^{(c)}(\BMx) \partder{\bar{S}_c   }{\pi_a}
                 +B_{k}^{(cd)}(\BMx)\partder{\bar{N}_{cd}}{\pi_a}
           \right] 
\Ea
\right.
\nome{3.14}
\Ee
(where the symbol $\deq$ means ``evaluated on the solutions of
the equations of motion'').

  Since, in phase space, the generators $\bar{S}_a$, $\bar{N}_{ab}$
have still the form they have in 
eqs.\EqRef{2.13},\EqRef{2.14}, because of  the second of eqs\EqRef{3.14},
we can write 
\Be
\Xi^{ab} (\BMx) \pi_b = \mu D r^a ~~,
\nome{3.15}
\Ee
where
\Be
\Xi^{ab} (\BMx) = \delta^{ab} 
          -\sqrt{\frac{\mu}{k}} \dot{x}^k B_{k}^{(ab)}(\BMx)
\;\; .\nome{3.16}
\Ee
Therefore, we obtain
\Be
\pi_a = \mu |\Xi^{-1}|_{ab} (\BMx) D r^b ~~.
\nome{3.17}
\Ee
Then, the first of eqs.\EqRef{3.14} gives
\Be
\Ba{rcl}
 P_i &=& M \delta_{ij} \dot{x}^j 
        + B_i^{(a)}  (\BMx) \bar{S}_a
        + B_i^{(ab)} (\BMx) \bar{N}_{ab} ~~,
\Ea
\nome{3.18}
\Ee
and we obtain the Lagrangian as:
\Be
\fl
\Ba{rcl}
\Lc_{SU(3)\otimes U(1)} &=& P_k \dot{x}^k + \pi_a \dot{r}^a - \bar{H}_{SU(3)\otimes U(1)} 
\\[2 mm]
      &=&\Ds \frac{M}{2} \delta_{ij} \dot{x}^i\dot{x}^j
           + \frac{\mu}{2} |\Xi^{-1}|_{ab} Dr^a ~ Dr^b
           - \frac{k}{2} \Xi^{ab} \delta_{ac}\delta_{bd} r^c r^d
~~.
\Ea
\nome{3.19}
\Ee

Finally, using eq.\EqRef{2.14}, we get
\Be
  B_k^{(ab)} (\BMx) \bar{N}_{ab}
 =  B_k^{\prime(ab)} (\BMx) \bar{\Qc}_{ab}
   + B_k^{\prime} (\BMx) \frac{\Ec^\prime}{3}
~~,
\Ee
and
\Be
\left\{
\Ba{rcl}
  B_k^{\prime (ab)}   (\BMx) 
&=& B_k^{(ab)} (\BMx) -\frac{1}{3} \delta^{ab}\delta_{cd}
                        B_k^{(cd)} (\BMx) 
\\[2 mm] 
  B_k^{\prime} (\BMx) &=& \delta_{cd} B_k^{(cd)} (\BMx)
~~.
\Ea
\right.
\nome{3.20}
\Ee
Eqs.\EqRef{3.20} identify the Yang-Mills fields, associated to the
{\it quadrupole moment} and to the {\it internal energy}, 
respectively, which contribute to
the potential energy  term through the symmetric 
tensor $\Xi^{ab} (\BMx)$.
Its inverse $|\Xi^{-1}|_{ab}$  behaves as a dynamically 
induced  ``internal metric'' 
for the scalar product of the contravariant vector $D\BMr=(Dr^a)$.
On the other hand, the SU(2) Yang-Mills fields determine
the ``internal spin connection'' 
$(^{\SSs Spin}\Gamma_{k~a}^b = B^{(c)}_k c_{ca}^{b})$
 and its associated ``internal  covariant derivative'' $Dr^a$.

  The $SU(3)\otimes U(1)$ generators, expressed in 
configuration space, become:
\Be
\fl
\left\{
\Ba{rcl}
\Ec^\prime  &=& \Ds  \frac{1}{2} \left( 
               \frac{\mu^2}{\sqrt{\mu k}}  
         \delta^{ab} |\Xi^{-1}|_{ac} (\BMx)
                     |\Xi^{-1}|_{bd} (\BMx) D r^c Dr^d 
               + \sqrt{\mu k}  \delta_{ab} r^a r^b \right) \\[2 mm]
S_{a}       &=& \Ds \mu c_{ab}^{~~c} r^b 
                |\Xi^{-1}|_{cd}  (\BMx) Dr^{d} \\[2 mm]
\Qc_{ab}    &=& \Ds \left[
         \frac{\mu^2}{2 \sqrt{\mu k}} 
          |\Xi^{-1}|_{ac}  (\BMx) D r^c 
          |\Xi^{-1}|_{bd}  (\BMx) D r^d
        +\frac{\sqrt{\mu k}}{2} \delta_{ac}\delta_{bd}  r^c r^d 
           \right] \\[2 mm]
  & & ~~ \Ds -\delta_{ab}\frac{\Ec^\prime}{3} 
\; .
\Ea
\right.
\nome{3.21}
\Ee
Now, under the $SU(3)\otimes U(1)$ transformations defined by (\ref{3.7})
and:
\Be
 \delta r^a = \Ds c_{bc}^{~~c} \theta^b(\BMx)~  r^c
              + \frac{\mu}{\sqrt{k}} \xi^{ab}(\BMx)~   
                |\Xi^{-1}|_{bc} (\BMx) D r^c ~~,
\nome{3.22}
\Ee
we get
\begin{eqnarray}
\fl
\delta \Lc_{SU(3)\otimes U(1)} = 
\nome{3.23}\\
\lo= \Ds \frac{d}{dt} \left[
          \sqrt{\frac{\mu}{k}} \xi^{ab}(\BMx)  
          \left(  \frac{\mu}{2}
                  ~|\Xi^{-1}|_{ac}  (\BMx) ~D r^c 
                  ~|\Xi^{-1}|_{bd}  (\BMx) ~D r^d
                -\frac{k}{2} \delta_{ac}\delta_{bd} r^c r^d
          \right) \right] ~~.
\nonumber
\end{eqnarray}
We want to impose now invariance ({\it quasi-invariance}) of the theory
with respect to local {\it gauge} variations dependent
on {\it time}, besides the {\it center-of-mass} coordinates,
i.e. for transformations of the form:
\Be
 \delta r^a = \Ds c_{bc}^{~~a} \theta^b(\BMx,t)~  r^c
              + \sqrt{\frac{\mu}{k}} \xi^{ab}(\BMx,t)~   
                |\Xi^{-1}|_{bc} (\BMx,t) D r^c ~~.
\nome{3.25}
\Ee
This task cannot be dealt with
in the non reparametrization-invariant Hamiltonian picture. 
It is easy to verify, however, that the only needed 
modification is the introduction of additional 
0-components for the $SU(3)\otimes U(1)$ fields
$B_i^{(a)}$ and $B_i^{(ab)}$, along with a modification of
the  transformation properties (\ref{3.7}), in the form
\Be
\fl
\left\{
\Ba{rcl}
  \delta B_{k}^{(a)}(\BMx,t)
             &=& \Ds  - \partder{\theta^a (\BMx,t)}{x^k}  
                      + c_{bc}^{~~a} \theta^b (\BMx,t) B_{k}^{(c)}(\BMx,t) 
\\[2 mm]
             & & \Ds  
                      + \xi^{cd} (\BMx,t)
                    [c_{ce}^{~~a} \delta_{df} B_{k}^{(ef)}(\BMx,t)
                    +c_{de}^{~~a} \delta_{cf} B_{k}^{(ef)}(\BMx,t)  ]  
\\[3 mm]
  \delta B_{0}^{(a)}(\BMx,t)
             &=& \Ds  - \partder{\theta^a (\BMx,t)}{t}  
                      + c_{bc}^{~~a} \theta^b (\BMx,t) B_{0}^{(c)}(\BMx,t) 
\\[2 mm]
             & & \Ds  
                      + \xi^{cd} (\BMx,t)
                    [c_{ce}^{~~a} \delta_{df} B_{0}^{(ef)}(\BMx,t)
                    +c_{de}^{~~a} \delta_{cf} B_{0}^{(ef)}(\BMx,t)  ]  
\\[3 mm]
  \delta B_{k}^{(ab)}(\BMx,t)
           &=&\theta^c (\BMx,t) [ c_{cd}^{~~a} B_{k}^{(db)} (\BMx,t)
                               +c_{cd}^{~~b} B_{k}^{(ad)} (\BMx,t) ]  
\\[2 mm]
           & &\Ds   + \frac{1}{4} \xi^{cd}(\BMx,t)
                 [c_{ec}^{~~a} \delta_d^b
                 +c_{ec}^{~~b} \delta_d^a
                 +c_{ed}^{~~a} \delta_c^b
                 +c_{ed}^{~~b} \delta_c^a] B_{k}^{(e)}(\BMx,t) 
\\[2 mm]
           & &\Ds     - \partder{\xi^{ab}(\BMx,t)}{x^k}
\\[3 mm]
  \delta B_{0}^{(ab)}(\BMx,t)
           &=&\theta^c (\BMx,t) [ c_{cd}^{~~a} B_{0}^{(db)} (\BMx,t)
                               +c_{cd}^{~~b} B_{0}^{(ad)} (\BMx,t) ]  
\\[2 mm]
           & &\Ds   + \frac{1}{4} \xi^{cd}(\BMx,t)
                 [c_{ec}^{~~a} \delta_d^b
                 +c_{ec}^{~~b} \delta_d^a
                 +c_{ed}^{~~a} \delta_c^b
                 +c_{ed}^{~~b} \delta_c^a] B_{0}^{(e)}(\BMx,t) 
\\[2 mm]
           & &\Ds     - \partder{\xi^{ab}(\BMx,t)}{t} ~~,
\Ea
\right.
\nome{3.26}
\Ee
and a redefinition of $Dr^a$ and $\Xi^{ab} (\BMx)$
to take into account the additional time components:
\Be
\Ba{rcl}
 D r^a &=& \Ds \dot{r}^a + \dot{x}^k B_{k}^{(b)} (\BMx,t) c_{bc}^{~~a} r^c
                     +  B_{0}^{(b)} (\BMx,t) c_{bc}^{~~a} r^c
\\[2 mm]
\Xi^{ab} (\BMx,t) &=& \Ds \delta^{ab} 
                    -\sqrt{\frac{\mu}{k}} 
         \left[ \dot{x}^k B_{k}^{(ab)}(\BMx,t)
               - B_{0}^{(ab)}(\BMx,t) \right] ~~.
\Ea
\nome{3.27}
\Ee
Then, we obtain the Lagrangian:
\Be
\fl
\Lc_{SU(3)\otimes U(1)} = \Ds \frac{M}{2} \delta_{ij} \dot{x}^i\dot{x}^j
      + \frac{\mu}{2} |\Xi^{-1}|_{ab}(\BMx,t) ~ Dr^a ~ Dr^b
      - \frac{k}{2} \Xi^{ab}(\BMx,t)~ \delta_{ac}\delta_{bd} r^c r^d ~~,
\nome{3.28}
\Ee
which is {\it quasi-invariant} under the transformation (\ref{3.25}).
In fact, we have:
\begin{eqnarray}
\fl
\delta \Lc_{SU(3)\otimes U(1)} = 
\nome{3.29}\\
\fl ~~= \Ds \frac{d}{dt} \left[
          \sqrt{\frac{\mu}{k}} \xi^{ab}(\BMx,t)  
          \left(  \frac{\mu}{2}
                  |\Xi^{-1}|_{ac}  (\BMx,t) ~D r^c 
                  |\Xi^{-1}|_{bd}  (\BMx,t) ~D r^d
                -\frac{k}{2} \delta_{ac}\delta_{bd} r^c r^d
          \right) \right]
~~.
\nonumber
\end{eqnarray}
It is worth stressing the peculiar role played by the
``internal'' covariant derivative and the ``internal'' metric in
the Lagrangian expression (\ref{3.28}).

Within the Hamiltonian picture, the generalization above
corresponds to the substitution rule
\Be
P_i \mapsto \Ds \Pc_i =  P_i 
                            - B_{i}^{(a)}(\BMx,t)  S_a
                            - B_{i}^{(ab)}(\BMx,t) N_{ab} 
~~,
\nome{3.30}
\Ee
and, consequently, to
\Be
H   \mapsto \Hc   = H[\Pc] - B_{0}^{(a)}(\BMx,t)  S_a
                           - B_{0}^{(ab)}(\BMx,t) N_{ab} ~~.
\Ee 
Let us remark that the {\it trace} parts 
$(B_{k}^{\prime}(\BMx,t),B_{0}^{\prime}(\BMx,t))$
of the $B$-fields transform separately as 
\Be
\Ba{rcl}
  \delta B_{k}^{\prime}(\BMx,t)
     &=&\Ds  - \partder{}{x^k}\left[\delta_{ab}\xi^{ab}(\BMx,t)\right]
\\[2 mm]
  \delta B_{0}^{\prime}(\BMx,t)
     &=&\Ds  - \partder{}{t}\left[\delta_{ab}\xi^{ab}(\BMx,t)\right]
~~~,
\Ea
\Ee
as indeed it must be, due to the Abelian character of the U(1) 
subgroup generated by the internal energy $\Ec^\prime$, to
which these fields are associated.

Following a similar way of deduction, is it possible
to find out the time variation law for the {\it conserved charge}.
Let $(W_i)_u$ and  $q_i$ be the gauge-fields and the position
variables, respectively, and let us denote their 
variation under the $\delta$ transformations by
\begin{eqnarray*}
  \delta q_i &=& \epsilon\; \phi_i[q,\dot{q}] \\
  \delta (W_i)_u &=& \epsilon\; \Phi_{i,u} [(W_i)_u ] ~~.
\end{eqnarray*}
Moreover, let us impose that the variation of the
Action under the transformation of these quantities 
be a total time derivative\footnote{As is well known, 
from this hypothesis, it follows 
that  $\delta$ is a symmetry of the system}\rlap :
\begin{equation}
 \delta S[q,\dot{q},(W_i)_u ]  = \int dt ~\delta {\cal L} 
            =  \int dt \frac{d}{dt} [\epsilon F]   \; .
\nome{S1}
\end{equation}
Finally, let us rewrite this variation of the Lagrangian in the form:
\begin{equation}
\fl
\delta S[q,\dot{q},(W_i)_u ]  = \int dt \left\{ {\left[ {   
    \partder{{\cal L}}{q_i} -  \frac{d}{dt} \partder{{\cal L}}{\dot{q}_i}
            } \right] \epsilon \phi_i   
    +  \frac{d}{dt}  \left[ { 
       \partder{{\cal L}}{\dot{q}_i}  \epsilon \phi_i   
            } \right]    
    +  \partder{{\cal L}}{(W_i)_u} \epsilon \Phi_{i,u} }\right\}
\; .
\nome{S2}
\end{equation} 
Then, since the expressions (\ref{S1}) and (\ref{S2}) 
must be equal, 
putting $\epsilon$ equal to constant and taking into account
the Euler-Lagrange equations 
$\partder{{\cal L}}{q_i} -  \frac{d}{dt} \partder{{\cal L}}{\dot{q}_i} =0$, 
it follows
\begin{equation}
 \frac{d}{dt}   \left[ { 
      F -   \partder{{\cal L}}{\dot{q}_i} \phi_i   
            } \right]
    =   \partder{{\cal L}}{(W_i)_u} \Phi_{i,u}   
\; .
\end{equation}
Note, in particular, that, in the case of the oscillator, the 
time variations of the standard generators of 
$SU(3)\otimes U(1)$, corresponding
to this way of  describing the transformations, are 
\begin{eqnarray*}
\frac{d}{dt} S_a 
&=&  \Ds c_{ab}^{~~c} {S}_c 
      \left( \dot{x}^i B_i^{(b)} + B_0^{(b)} \right)
       +2 c_{ac}^{~~b} {N}_{bd}  
      \left( \dot{x}^i B_i^{(cd)} + B_0^{(cd)} \right)
\\[2 mm]
\frac{d}{dt} N_{ab} 
&=&\Ds\left(  c_{ac}^{~~d} \bar{N}_{db}
            + c_{bc}^{~~d} \bar{N}_{ad}
      \right) \left( \dot{x}^i B_i^{(c)} + B_0^{(c)}
      \right) \\[2 mm]
& &~+\frac{1}{2} \left( 
       c_{ac}^{~~e} \delta_{bd} 
      +c_{bc}^{~~e} \delta_{ad} \right) \bar{S}_{e}
      \left( \dot{x}^i B_i^{(cd)} + B_0^{(cd)}
      \right)
\; .
\end{eqnarray*}


\section{Gauging the extended Galilei Group for the free particle}

In reference \cite{DePietri95} we have shown that by 
{\it gauging {\sl \`a la}} Utiyama \cite{Uty} the realization 
of the {\it extended} Galilei group associated 
to a non-relativistic free particle of mass $M$,
one is led to the Lagrangian:
\Be
{\Lc}_M^{g} (t) = {1 \over \Theta} \frac{M}{2}
                           \left[{   g_{ij} \dot{x}^i  \dot{x}^j
                                    + 2 A_i \dot{x}^i
                                    + 2 A_0
                            } \right]   
~~.
\nome{LGG}
\Ee
This Lagrangian is {\it quasi-invariant} under the mass-point
coordinate transformations
\Be
\left\{
\Ba{lcl}
{\delta}_0 ~t  &=& 0 \\
{\delta}_0 x^i  &=& \varepsilon (t) \dot{x}^i
                    +\eta^i ({\BMx},t) - t~v^i({\BMx},t)   \\[2 mm]
{\delta}_0 \dot{x}^i  
         &=& \Ds \frac{d}{dt} [ \varepsilon (t) \dot{x}^i ]
                      +\dot{x}^k\partder{}{x^k}
                        [\eta^i({\BMx},t)-t~v^i({\BMx},t)]  \\[2 mm]
         & & \Ds + \partder{}{t}[\eta^i ({\BMx},t) - t~v^i({\BMx},t)] 
~~~,
\Ea
\right.
\nome{3.6I}
\Ee
and the field transformations
\Be
\left\{
\Ba{rcl}
{\delta}_{0} \Theta    &=& \Ds  \hat{\delta}_0 \Theta
                        = \dot{\varepsilon} \Theta \\[2 mm]
{\delta}_{0}   A_0     &=& \Ds \hat{\delta}_0 A_0
                         - {\varepsilon} \partder{A_0}{t}
                         + A_{i,j} \tilde\eta^j  \\[2 mm]
                     &=& \Ds 2 \dot{\varepsilon} A_0
                         - A_i \partder{\tilde\eta^i}{t}
                         + \Theta \partder{\Fc}{t} \\[2 mm]
{\delta}_{0}   A_i     &=& \Ds  \hat{\delta}_0 A_i
                        - {\varepsilon} \partder{A_i}{t}
                        + A_{i,j} \tilde\eta^j \\[2 mm]
                     &=& \Ds \dot{\varepsilon} A_i
                        - A_j \partder{\tilde\eta^j}{x^i}
                        - g_{ij}  \partder{\tilde\eta^j}{t}
                        + \Theta \partder{\Fc}{x^i} ~~,
\Ea
\right.
\nome{FTR}
\Ee 
in the sense that
\Be
{\delta}_{0} {\Lc}_M^{g} =   {d \Fc \over dt} 
                       + \varepsilon {d [ {\Lc}_M^{g} ] \over dt}  
\Ee
holds, so that 
\Be
\Delta {\Sc}_M^{g} = \int_{t_1}^{t_2} dt  \left[ { {d \Fc \over dt} 
                         +{d [ \varepsilon  {\Lc}_M^{g} ] \over dt}
                                  } \right] = 0  ~~,
\Ee
where $\Fc= M g_{ij} v^i(x,t) x^j $ is the {\it cocycle} term connected
to the {\it central extension} of the group.

For later use it is instrumental to report explicitly 
the deduction of this result in the 
reparameterization-invariant Hamiltonian  description.
Let $\BMx(\tau)$, $t(\tau)$, be the space and time coordinates of the
free mass-point. The Lagrangian and the Action can be written
\Be
\Ba{rl}
\hat{\Lc}_M (\lambda) &= \Ds \frac{1}{2} M {\delta_{ij} \primato{x}^i(\lambda)
          \primato{x}^j(\lambda) \over \primato{t}(\lambda)} 
          ~~~, \\[4 mm]
\hat{\Ac}_M &= \Ds \int_{{\lambda}_1}^{{\lambda}_2}
          d{\lambda}~~\hat{\Lc}_M(\lambda) ~~~,
\Ea
\nome{4.19}
\Ee
where  $\primato{f}(\lambda ) \equiv \frac{d}{d{\lambda}} f(\lambda )$.
The canonical momenta are
\Be
\left\{ {
\Ba{rl}
  \hat{E} &= \Ds -\partder{\hat{\Lc}}{\primato{t}}
           = \frac{M}{2} {\delta_{ij} \primato{x}^i(\lambda)
               \primato{x}^j(\lambda) \over \primato{t}^2(\lambda)}
           = \frac{1}{2M} \delta^{ij} \hat{p}_i \hat{p}_j \\[4 mm]
   \hat{p}_i &= \Ds \partder{\hat{\Lc}}{\primato{x}^i}
           = M {\delta_{ij} \primato{x}^j(\lambda)
                   \over \primato{t}(\lambda)}
           = M \delta_{ij} \dot{x}^j(t) = p_i  ~~~,
\Ea
}\right.
\nome{4.20}
\Ee
where we have defined the Poisson brackets so that:
\Be
\Ba{rl}
         \{ t(\lambda ) , \hat{E}(\lambda ) \}^\prime &= -1 \\[2 mm]
         \{ x^i(\lambda ),\hat{p}_j(\lambda )\}^\prime &= \delta^i_j ~~.
\Ea
\nome{4.21}
\Ee
In the enlarged phase-space, coordinatized by $t,~x^i,~\hat{E},~\hat{p}_i$,
we get a vanishing canonical Hamiltonian and the first-class constraint
\Be
\hat{\chi} \equiv \hat{E} - \frac{1}{2M} \delta^{ij} \hat{p}_i \hat{p}_j
           \approx 0 ~~.
\nome{4.22}
\Ee
The constraint $\hat{\chi}$ generates the following transformations
of the configurational variables:
$\delta t(\lambda) = - \alpha (\lambda ) ~,~~
 \delta x^i (\lambda) = - \alpha (\lambda ) \frac{1}{M} \delta^{ij} \hat{p}_j
$.
This corresponds to the reparameterization {\it gauge} transformation
$\lambda \rightarrow \lambda + \alpha(\lambda ) / t^\prime$.
The Lagrangian is obviously {\it quasi-invariant} under this 
operation since
\Be
\delta\hat{\Lc}_M=\frac{d}{d{\lambda}}
                \left[ {-\alpha (\lambda ) \over t^\prime}
                        \hat{\Lc}_M
                \right]  ~.
\nome{4.23}
\Ee
The canonical generators of the {\it extended} Galilei 
group are now:
\Be
\fl
\hat{\Hc} = \hat{E}~, ~~\hat{\Pc}_i = \hat{p}_i~,
               ~~\hat{\Kc}_i = M~\delta_{ij}~x^j - t~\hat{p}_i ~,
               ~~\hat{\Jc}_i = c^{~~k}_{ij} x^j \hat{p}_k ~,
               ~~\hat{\Mc}   = M ~,
\nome{4.24}
\Ee
and satisfy the Lie-algebra (\ref{2.8}) with 
the primed Poisson-brackets (\ref{4.21}).
Consequently, the generator of the complete 
phase-space Galilei transformation
$\hat{\bar{\delta}}$~, which is now given by
\Be
\Ba{rl}
 \hat{\bar{G}} &= \varepsilon \hat{E}  + \varepsilon^i \hat{p}_i
                     + \omega^i \hat{\bar{J}}_i + v^i \hat{\bar{K}}_i \\
         &= \varepsilon \hat{E} + (\eta^i-t v^i) \hat{p}_i
                     + M \delta_{ij}  v^i x^j     ~~,\\
\Ea
\nome{4.25}
\Ee
yields the following ``equal-$\lambda$'' infinitesimal transformations:
\Be
\left\{
\Ba{ccl}
\hat{\bar\delta} \lambda &=& 0 \\
\hat{\bar\delta} t   &=& \{ t~ ,\hat{\bar{G}}\}
                      = - \varepsilon = \hat\delta_0 t = \delta t  \\
\hat{\bar\delta} x^i &=& \{ x^i,\hat{\bar{G}}\}
                      = \varepsilon^i+c^{~~i}_{jk}\omega^j x^k -t v^i
                      = \hat\delta_0 x^i  = \delta x^i
\;\; .
\Ea
\right.
\nome{4.26}
\Ee
We have now
\mbox{
$\left.{ \hat\delta \hat{p}_i } \right|_{\hat{p} = 
\partial \hat{\Lc}_M /
\partial \primato{x} }
  = \hat\delta \left[{M \delta_{ij} \primato{x}^j \over \primato{t} }
                 \right]
$},
without any use of Euler-Lagrange equations.
On the other hand, under the transformations (\ref{4.26}), it follows:
\Be
\hat{\bar{\delta}} \hat{\Lc}_M  = \frac{d}{d{\lambda}} \left[{
                                  {- M \delta_{ij} v^i x^j } }\right]
 ~~~,
\nome{4.27} 
\Ee
and
\Be
\hat{\bar{\delta}} \hat{\chi} = 0 ~~, 
\nome{4.28} 
\Ee
so that the canonical generators (\ref{4.24}) are constants
of the motion, as they must be. Moreover, the ({\it first-class}) 
constraint (\ref{4.22}) is Galilei
invariant, and the {\it quasi-invariance} of the Lagrangian is an
effect of the {\it central-charge} term alone.

We want now to impose the condition that the theory be invariant
under the {\it localized} Galilei transformation
generated by:
\Be
\Ba{rl}
\hat{\bar{G}} &= \varepsilon (t)~\hat{E}
                     + \varepsilon^i({\BMx},t)~ \hat{p}_i
                     + \omega^i({\BMx},t)~ \hat{\bar{J}}_i
                     + v^i ({\BMx},t)~\hat{\bar{K}}_i \\[1 mm]
                  &= \varepsilon (t)~\hat{E}
                     + [\eta^i({\BMx},t)~-t v^i({\BMx},t)]~\hat{p}_i
                     + M~ \delta_{ij}  v^i ({\BMx},t)~x^j ~~,
\Ea
\nome{eqNG}
\Ee
i.e.:
\Be
\fl
\left\{
\Ba{lcl}
\hat{\bar\delta}_0 ~t(\lambda )  
&=& -\varepsilon (t(\lambda ))   \\
\hat{\bar\delta}_0 x^i(\lambda )
&=&  \eta^i ({\BMx},t) - t~v^i({\BMx},t) \\[2 mm]
\hat{\bar\delta}_0 \hat{p}_i  &=& \Ds -\hat{p}_k\partder{}{x^i}
                             [\eta^k({\BMx},t)-t~v^k({\BMx},t)]
                             -M \partder{}{x^i}
                             [\delta_{lk}x^l v^k({\BMx},t)]\\[2 mm]
\hat{\bar\delta}_0 \hat{E}    &=& \Ds \hat{E} \frac{d\varepsilon(t)}{dt}
                             +\hat{p}_i \partder{}{t}
                             [\eta^i({\BMx},t)-t~v^i({\BMx},t)]
                             + M \partder{}{t}
                             [\delta_{ij}x^i v^j({\BMx},t)] ~~.
\Ea
\right.
\nome{LocalG}
\Ee
Let us consider first the phase-space approach. By introducing
the minimal coupling (see Ref.\cite{DePietri95}, Section 3):
\Be
\left\{
\Ba{rcl}
P_i &\mapsto& \Ds \Pc_a = {\bf H}_{a}^k(\BMx,t) 
                      ( P_k - \frac{M}{\Theta} A_k(\BMx,t)  ) 
\\[2 mm]
E &\mapsto& \Ds \Ec  = \frac{1}{\Theta} 
                      ( E - \frac{M}{\Theta} A_0(\BMx,t)  ) ~~,
\Ea
\right.
\nome{4.29}
\Ee
we get the following {\it gauge} constraint:
\Be
\hat{\chi}_g =  {1 \over \Theta} [\hat{E} + {M \over \Theta} A_0 ]
           -  {g^{ij} \over 2 m} [\hat{p}_i - {M \over \Theta} A_i ]
                                 [\hat{p}_j - {M \over \Theta} A_j ]
                \simeq 0 ~~.
\nome{4.30}
\Ee
Note that the counterpart of the second equation (\ref{4.29})
in the standard (non reparameterization-invariant) picture
is:
\Be
H   \mapsto \Hc   = \Theta(t) H[\Pc] - \frac{M}{\Theta} A_0(\BMx,t)  ~~.
\Ee 
Since, under the transformation (\ref{LocalG}), we have
\Be
\Ba{rl}
\hat\delta_0 \hat\chi 
    &=\Ds {1 \over \Theta} \hat{\bar\delta} \hat{E}
         -{1 \over \Theta^2} \left[{
                   \hat{E} + {M \over \Theta} A_0 }\right]
                   \hat\delta_0\Theta   
         + {1 \over \Theta^2} M  \hat\delta_0 A_0   \\[1 mm]
    &~\Ds - {1 \over 2 M} \left( \hat\delta_0 g^{ij} \right)
                   [\hat{p}_i - {M \over \Theta} A_i ]
                   [\hat{p}_j - {M \over \Theta} A_j ]  
\\[1 mm]
    &~\Ds - {g^{ij} \over M}  
            [\hat{p}_i - {M \over \Theta} A_i ]
            [\hat{\bar\delta}\hat{p}_j
          - {M \over \Theta}  \hat\delta_0 A_j
          + {M \over \Theta^2} \hat\delta_0\Theta A_j ] ~~,
\Ea
\nome{4.31}
\Ee
the invariance of the constraint
\Be
\hat\delta_0 \hat\chi_g = 0 ~~,
\Ee
is satisfied provided that the fields  transform
according to
\Be
\left\{
\Ba{rl}
\hat{\delta}_0 \Theta  &= \dot{\epsilon}(t) \Theta (t)  \\[2 mm]
\hat{\delta}_0  g_{ij} &= \Ds - \partder{\tilde\eta^k ({\bf x},t) }{x^i} g_{kj}
                        - \partder{\tilde\eta^k ({\bf x},t) }{x^i} g_{kj} \\[2 mm]
\hat{\delta}_0 A_0     &= \Ds 2 \dot{\varepsilon} A_0
                        - A_i \partder{\tilde\eta^i}{t}
                        - \Theta \partder{}{t}
                                 \left[{g_{ij} v^i x^j}\right] \\[2 mm]
\hat{\delta}_0 A_i     &= \Ds \dot{\varepsilon} A_i
                        - A_j \partder{\tilde\eta^j}{x^i}
                        - g_{ij}  \partder{\tilde\eta^j}{t}
                        - \Theta \partder{}{x^i}
                                 \left[{g_{ij} v^i x^j}\right]  ~~.
\Ea
\right.
\nome{4.38}
\Ee
The {\it gauge} reparameterization-invariant Lagrangian 
corresponding to the constraints (\ref{4.30}) is:
\Be
\hat{\Lc}_M^{g} (\lambda)
                    = {1\over\Theta\primato{t}} \frac{M}{2}
                           \left[ {  g_{ij} \primato{x}^i  \primato{x}^j
                                    + 2 A_i \primato{x}^i  \primato{t}
                                    + 2 A_0 \primato{t}    \primato{t} }
                           \right] ~,
\nome{4.39}
\Ee
with
\Be
\hat{\delta} \hat{\Lc}_M^{g} = {d \Fc \over d\lambda} ~~,
\nome{4.40}
\Ee
and it  corresponds indeed to the ``standard'' Action (\ref{LGG}).
It is worth remarking that, while in the standard
Lagrangian picture the specific form of the 
{\it cocycle} term $\Fc$ must be guessed, the  latter
is {\it explicitly determined} within the Hamiltonian 
picture. In fact, since the algebraic properties are 
essential in the Hamiltonian
formalism, the {\it cocycle} term is a necessary consequence 
of the {\it central extension} of the group.

Note finally that, were it not for the presence of the
{\it einbein-like} component $\Theta$, the modified Hamiltonian
constraint (\ref{4.22}) could have been made invariant only in
the weak sense $\hat\delta \hat\chi_g = \varepsilon \hat\chi_g
\approx 0$.


\section{Gauging the maximal symmetry group in the one-time theory}

We are now ready to {\it gauge} the Galilei kinematical 
{\it plus} the internal dynamical group
of the harmonic oscillator in a {\it unified way}. 
As shown in Section {\bf 1}, the algebra of the
maximal dynamical symmetry  of the harmonic
oscillator is the semi-direct sum of
the {\it kinematical extended} Galilei algebra times  
the {\it dynamical} $SU(3)\otimes U(1)$ algebra
given by (\ref{2.8},\ref{2.15},\ref{2.19}). 
We proceed by letting the infinitesimal
parameters of the transformations (\ref{2.11}) and
(\ref{2.16}) to become dependent on the {\it center-of-mass} coordinates
and {\it time}; then, by searching for a new Action invariant ({\it
quasi-invariant}) under the transformations
\Be
\fl
\left\{
\Ba{rcl}
 \delta_0 x^i &=&\Ds \epsilon(t) \dot{x}^i
                  +\epsilon^i(\BMx,t) 
                  + c_{jk}^{~~l} \omega^j(\BMx,t) x^k
                  - t v^i(\BMx,t) \\
              &=&\Ds \epsilon(t) \dot{x}^i
                  +\tilde{\eta}^i(\BMx,t)  \\
 \delta_0 r^a &=&\Ds \epsilon(t) \dot{r}^a 
                  + c_{bc}^{~~a} \omega^b(\BMx,t) r^c
                  + c_{bc}^{~~a} \theta^b(\BMx,t) r^c
                  + \sqrt{\frac{\mu}{k}} \xi^{ab}(\BMx,t) 
                    \delta_{bc} \dot{r}^c \\
              &=&\Ds \epsilon(t) \dot{r}^a 
                  + c_{bc}^{~~a} \tilde{\theta}^b(\BMx,t) r^c
                  + \sqrt{\frac{\mu}{k}} \xi^{ab}(\BMx,t) 
                    \delta_{bc} \dot{r}^c ~~,
\Ea
\right.
\nome{xxx}
\Ee
(where we have set  $\tilde{\eta}^i(\BMx,t) \equiv
\epsilon^i(\BMx,t) + c_{jk}^{~~l} \omega^j(\BMx,t) x^k - t
v^i(\BMx,t)$ and $\tilde{\theta}^b(\BMx,t)= {\theta}^b(\BMx,t) +
\omega^b(\BMx,t)$).    
Instructed by the results obtained in Section {\bf 3}, we know that
the compensating $SU(3)\otimes U(1)$ Yang-Mills fields have to appear in the 
``internal'' covariant derivative and in the ``internal'' metric.
Furthermore, the {\it dynamical} component of the $SU(3)\otimes U(1)$ symmetry
must be modified. Finally, from Section {\bf 4}, we  know
in addition that, in {\it gauging} the Galilei part, 
a field  $\Theta(t)$ must necessarily be introduced
to take into account the redefinition of absolute time.

As a matter of fact, the required generalization can be achieved 
in the following way: 

\noindent
1) define the ``internal'' covariant derivative
and ``internal'' metric as:
\Be
\fl
\left\{
\Ba{rcl}
 D r^a &=& \Ds \dot{r}^a + \dot{x}^k B_{k}^{(b)} (\BMx,t) c_{bc}^{~~a} r^c
                     +  B_{0}^{(b)} (\BMx,t) c_{bc}^{~~a} r^c
\\[2 mm]
\Xi^{ab} (\BMx,t) &=& \Ds \delta^{ab} 
             -\frac{1}{\Theta(t)}\sqrt{\frac{\mu}{k}} 
         \left[ \dot{x}^k B_{k}^{(ab)}(\BMx,t)
               - B_{0}^{(ab)}(\BMx,t) \right] ~~~,
\Ea
\right.
\nome{5.1}
\Ee
2) define the dynamical modified symmetry transformations as:
\Be
\fl
\left\{
\Ba{rcl}
 \delta_0 x^i &=& \Ds \epsilon(t) \dot{x}^i + \tilde{\eta}^i(\BMx,t)  
\\[2 mm]
 \delta_0 r^a &=& \Ds \epsilon(t) \dot{r}^a 
                  + c_{bc}^{~~a} \tilde{\theta}^b(\BMx,t) r^c
                  + \frac{1}{\Theta}\sqrt{\frac{\mu}{k}} \xi^{ab}(\BMx,t) 
                    |\Xi^{-1}|_{bc} (\BMx,t) \dot{r}^c ~~.
\Ea
\right.
\nome{5.2}
\Ee
3) write the final complete Lagrangian in the form:
\Be
\fl
\Ba{rcl}
\Lc_{\Sss G} &=& \Ds \frac{M}{2\Theta(t)} 
                \left[ g_{ij}(\BMx,t)~  \dot{x}^i\dot{x}^j 
                      + 2 A_i(\BMx,t)~  \dot{x^i} 
                      + 2 A_0(\BMx,t)
                \right] \\[2 mm] 
      &&\Ds   + \frac{\mu}{2\Theta} |\Xi^{-1}|_{ab}(\BMx,t)~Dr^a~Dr^b
              -\frac{k\Theta}{2} \Xi^{ab}(\BMx,t)~ 
               \delta_{ac}\delta_{bd} r^c r^d ~~.
\Ea
\nome{5.3}
\Ee
Then, the complete transformation rules for the Yang-Mills
fields are derived by requiring that the Action defined by the
Lagrangian (5.4) be {\it quasi-invariant} under the maximal dynamical 
symmetry group {\it gauged} {\sl \`a la} Utiyama, 
precisely in the form:
\Be
\fl
\Ba{rcl}
\lefteqn{ \Ds \delta_0 \int dt ~\Lc_{\Sss G}
          =\Ds \int dt~\left[ \delta_0~\Lc_{\Sss G}
                           -\Lc_{\Sss G}\dot{\epsilon} 
                    \right] } \\[2 mm]
      &=& \Ds\int dt~ \frac{d}{dt} \left[
          - m g_{ij} (\BMx,t) v^i(\BMx,t)~x^j
          \right] \\[2 mm]   
      & &+\Ds\int dt~ \frac{d}{dt} \left[
          \sqrt{\frac{\mu}{k}} \xi^{ab}(\BMx,t)  
          \left(  \frac{\mu}{2\Theta}
                  |\Xi^{-1}|_{ac}  (\BMx,t) ~D r^c 
                  |\Xi^{-1}|_{bd}  (\BMx,t) ~D r^d 
               \right. \right. \\[2 mm]
       & & ~~~~~~~~~~~~\Ds \left. \left.
          -\frac{k\Theta}{2} \delta_{ac}\delta_{bd} r^c r^d
          \right) \right]  ~~.
\Ea
\nome{5.4}
\Ee
Clearly, the transformation rules of the {\it gauge} fields associated 
to the Galilei group and the dynamical symmetry $SU(3)\otimes U(1)$, 
when considered separately,   have
to be exactly  those  we have just found for the separate case of 
the {\it gauged} Galilei group (Section 4) and of the the 
{\it gauged} $SU(3)\otimes U(1)$ dynamical group (section 3).
However, given the semi-direct product structure of the overall symmetry,
the $SU(3)\otimes U(1)$ Yang-Mills fields must undergo 
{\it additional transformations} under the Galilei group.
 
In carrying out the calculation, it is essential to take into account
the fact that we are imposing invariance of the Action with 
respect to transformations depending on the evolution 
parameter explicitly. Therefore, as shown by eq.(\ref{2.7a}), we have
to consider the following relation between the pure
field variation and the actual variation imposed
on the Lagrangian:
$\delta_0 f(\BMx,t) = \delta f(\BMx,t) +
\frac{d}{dt}f(\BMx,t) ~\dot{\epsilon}(t)$.
Then, it follows that the compensating {\it gauge} fields 
corresponding to variations (\ref{5.2}) satisfy the following 
transformation rules:
\Bea
\fl \hat{\delta}_0^g \Theta (t)  &=& \dot{\epsilon}(t) \Theta (t)  
\nonumber \\[2 mm] \fl
\hat{\delta}_0^g  g_{ij} (\BMx,t) &=& \Ds 
           - \partder{\tilde\eta^k ({\bf x},t) }{x^i} g_{kj}
           - \partder{\tilde\eta^k ({\bf x},t) }{x^i} g_{kj}   
\nonumber \\[2 mm] \fl
\hat{\delta}_0^g A_0 (\BMx,t)     &=& \Ds 2 \dot{\varepsilon} A_0
           - A_i \partder{\tilde\eta^i}{t}
           - \Theta \partder{}{t}
           \left[{g_{ij} v^i x^j}\right] 
\nonumber \\[2 mm] \fl
\hat{\delta}_0^g A_i (\BMx,t)     &=& \Ds \dot{\varepsilon} A_i
           - A_j \partder{\tilde\eta^j}{x^i}
           - g_{ij}  \partder{\tilde\eta^j}{t}
           - \Theta \partder{}{x^i}
           \left[{g_{jk} v^j x^k}\right]  ~~.
\nome{5.5}
\\[2 mm] \fl
\hat{\delta}_0^g B_{k}^{(a)}(\BMx,t)
             &=& \Ds   c_{bc}^{~~a} \tilde{\theta}^b (\BMx,t) 
                       B_{k}^{(c)}(\BMx,t)   
\nonumber \\[2 mm] \fl
             & & \Ds  + \xi^{cd} (\BMx,t)
                    [c_{ce}^{~~a} \delta_{df} B_{k}^{(ef)}(\BMx,t)
                    +c_{de}^{~~a} \delta_{cf} B_{k}^{(ef)}(\BMx,t)  ]  
\nonumber \\[2 mm] \fl
             & & \Ds
           - \partder{\tilde\eta^i}{x^k} B_{i}^{(a)}(\BMx,t) 
           - \partder{\tilde{\theta}^a (\BMx,t)}{x^k}  
\nonumber \\[3 mm] \fl
\hat{\delta}_0^g B_{0}^{(a)}(\BMx,t)
             &=& \Ds   c_{bc}^{~~a} \tilde{\theta}^b (\BMx,t) 
                       B_{0}^{(c)}(\BMx,t)
\nonumber \\[2 mm] \fl
             & & \Ds  + \xi^{cd} (\BMx,t)
                    [c_{ce}^{~~a} \delta_{df} B_{0}^{(ef)}(\BMx,t)
                    +c_{de}^{~~a} \delta_{cf} B_{0}^{(ef)}(\BMx,t)  ]  
\nonumber \\[2 mm] \fl
             & & \Ds 
           + \dot{\epsilon} B_{0}^{(a)}(\BMx,t) 
           - \partder{\tilde\eta^i}{t} B_{i}^{(a)}(\BMx,t) 
           - \partder{\tilde{\theta}^a (\BMx,t)}{t}  
\nonumber \\[3 mm] \fl
\hat{\delta}_0^g B_{k}^{(ab)}(\BMx,t)
           &=&\tilde{\theta}^c (\BMx,t) 
                 [ c_{cd}^{~~a} B_{k}^{(db)} (\BMx,t)
                  +c_{cd}^{~~b} B_{k}^{(da)} (\BMx,t) ]  
\nonumber \\[2 mm] \fl
           & &\Ds   + \frac{1}{4} \xi^{cd}(\BMx,t)
                 [c_{ec}^{~~a} \delta_d^b
                 +c_{ec}^{~~b} \delta_d^a
                 +c_{ed}^{~~a} \delta_c^b
                 +c_{ed}^{~~b} \delta_c^a] B_{k}^{(e)}(\BMx,t) 
\nonumber \\[2 mm] \fl
           & &\Ds
           - \partder{\tilde\eta^i}{x^k} B_{i}^{(ab)}(\BMx,t) 
           - \partder{\xi^{ab}(\BMx,t)}{x^k}
\nonumber \\[3 mm] \fl
\hat{\delta}_0^g B_{0}^{(ab)}(\BMx,t)
           &=&\tilde{\theta}^c (\BMx,t) 
                  [ c_{cd}^{~~a} B_{0}^{(db)} (\BMx,t)
                   +c_{cd}^{~~b} B_{0}^{(da)} (\BMx,t) ]  
\nonumber \\[2 mm] \fl
           & &\Ds   + \frac{1}{4} \xi^{cd}(\BMx,t)
                 [c_{ec}^{~~a} \delta_d^b
                 +c_{ec}^{~~b} \delta_d^a
                 +c_{ed}^{~~a} \delta_c^b
                 +c_{ed}^{~~b} \delta_c^a] B_{0}^{(e)}(\BMx,t) 
\nonumber \\[2 mm] \fl
           & &\Ds
           + \dot{\epsilon} B_{0}^{(ab)}(\BMx,t) 
           - \partder{\tilde\eta^i}{t} B_{i}^{(ab)}(\BMx,t) 
           - \partder{\xi^{ab}(\BMx,t)}{t} ~~.
\nonumber
\Eea
In conclusion, we see from equations \EqRef{5.1} and \EqRef{5.3}, 
that we have to take into account with the following {\it gauge} fields:
\begin{itemize}
\item[1)] either the {\it Euclidean metric} $g_{ij}(\BMx,t)$ 
          ($(g^{-1})^{ij} (\BMx,t)$) or the dreibein $\BME^a_i (\BMx,t)$
          ($\BMH_a^i (\BMx,t)$);
\item[2)] the time {\it reparameterization} field $\Theta(t)$;
\item[3)] the {\it inertial-gravitational} fields $A_i(\BMx,t)$ 
          and $A_0(\BMx,t)$;
\item[4)] the {\it ``spin connection''} fields $B^{(a)}_i(\BMx,t)$, 
          $B^{(a)}_0(\BMx,t)$, which generate
          the {\it internal covariant derivative};
\item[5)] the {\it quadrupole} and {\it internal energy} Yang-Mills fields
          $B^{(ab)}_i(\BMx,t)$, $B^{(ab)}_0(\BMx,t)$, 
          which generate the {\it internal metric}
          $\Xi^{rs}(\BMx,t)$,
\end{itemize}

Finally, let us remark that, going back to the Hamiltonian 
picture, we would find that the above procedure 
corresponds to  minimal substitution rules for energy and canonical 
momenta given by:
\Be
\Ba{rcl}
\fl
P_i &\mapsto& \Ds \Pc_a = {\bf H}_{a}^i(\BMx,t) 
                 \bigg[ P_i - \frac{M}{\Theta} A_i(\BMx,t)
                            - B_{i}^{(a)}(\BMx,t)  S_a
                            - B_{i}^{(ab)}(\BMx,t) N_{ab}  \bigg] 
\\[4 mm]
\fl
E &\mapsto& \Ds \Ec  = \frac{1}{\Theta} 
                  \bigg[ E - \frac{M}{\Theta} A_0(\BMx,t)
                            - B_{0}^{(a)}(\BMx,t)  S_a
                            - B_{0}^{(ab)}(\BMx,t) N_{ab}  \bigg] ~~,
\Ea
\nome{5.6}
\Ee
while, for the Hamiltonian itself, we would find:
\Be
\fl
H  \mapsto \Hc   = \Theta(t) H[\Pc] - \frac{M}{\Theta} A_0(\BMx,t) 
                                     - B_{0}^{(a)}(\BMx,t)  \bar{S}_a
                                     - B_{0}^{(ab)}(\BMx,t) \bar{N}_{ab} ~~.
\Ee 
How  all this can work will become clear presently in  approaching
the whole problem within the multi-temporal picture.

It is interesting to rewrite the final Lagrangian (\ref{5.3}) 
in terms of the  {\it individual} coordinates of the constituents
\Be
\fl
\Ba{rcl}
\Lc_{\Sss G} &=& 
  \Ds \frac{m_1}{2\Theta(t)} 
      \left[ \frac{m_1}{M}  g_{ij}(\BMx,t)
           +\frac{\mu}{m_1} |\Xi^{-1}|_{ij}(\BMx,t) 
      \right] \dot{x}^i_1 \dot{x}^j_1 \\[3mm] 
&&+\Ds \frac{m_2}{2\Theta(t)} 
      \left[ \frac{m_2}{M}  g_{ij}(\BMx,t)
           +\frac{\mu}{m_2} |\Xi^{-1}|_{ij}(\BMx,t) 
      \right] \dot{x}^i_2 \dot{x}^j_2 \\[3mm] 
&&+\Ds \frac{\mu}{\Theta(t)} 
      \left[ g_{ij}(\BMx,t)
           - |\Xi^{-1}|_{ij}(\BMx,t) 
      \right] \dot{x}^i_1 \dot{x}^j_2 \\[3mm] 
&&+\Ds \frac{m_1}{\Theta} \left[ \dot{x}_1^k A_k(\BMx_1,t) 
                 + A_0(\BMx_1,t) \right]
      +\frac{m_2}{\Theta} \left[ \dot{x}_2^k A_k(\BMx_2,t) 
                 + A_0(\BMx_2,t) \right] \\[3mm]
&&+\Ds \sum_{\alpha=1}^{\infty} \frac{\mu^\alpha}{\Theta}
       \left[ 
       \left(\frac{1}{m_2}\right)^\alpha
          \left( \dot{x}_2^k A_{k,i_1\cdots i_\alpha}(\BMx_2,t) 
           + A_{0,i_1\cdots i_\alpha}(\BMx_2,t) \right)
\right. 
\\[3mm]
      &&\Ds~~~~~~~~~~~~\left.
      +\left(\frac{-1}{m_1}\right)^\alpha
          \left( \dot{x}_1^k A_{k,i_1\cdots i_\alpha}(\BMx_1,t) 
           + A_{0,i_1\cdots i_\alpha}(\BMx_1,t) \right)
       \right] \cdot

\\[3mm]
      &&\Ds~~~~~~~~~ \cdot (x_1^{i_1}-x_2^{i_1})\cdots
(x_1^{i_\alpha}-x_2^{i_\alpha}) 
\\[3mm]
&&\Ds + \frac{\mu}{\Theta} |\Xi^{-1}|_{ab}(\BMx,t)~\dot{r}^a~
         \epsilon_{ef}^{~~b} 
         \left[\dot{x}^k B_{k}^{(e)} (\BMx,t) +  B_{0}^{(e)} (\BMx,t)
         \right] (x_1^{f}-x_2^{f}) \\[3mm]
&&\Ds + \frac{\mu}{2\Theta} |\Xi^{-1}|_{ab}(\BMx,t)
         \epsilon_{ef}^{~~a} 
         \left[\dot{x}^k B_{k}^{(e)} (\BMx,t) +  B_{0}^{(e)} (\BMx,t)
         \right] (x_1^{f}-x_2^{f}) \\[3mm]
&&\Ds~~~~~~~~~~~~~~~~~~~~
         ~\epsilon_{gh}^{~~b} 
         \left[\dot{x}^k B_{k}^{(g)} (\BMx,t) +  B_{0}^{(g)} (\BMx,t)
         \right] (x_1^{h}-x_2^{h}) \\[3mm]
&&\Ds       -\frac{k\Theta}{2} \Xi^{ab}(\BMx,t)~ 
       \delta_{ac}\delta_{bd} (x_1^{c}-x_2^{c})(x_1^{d}-x_2^{d})  ~~.
~~~,
\Ea
\nome{5.9}
\Ee
where the expressions of the fields calculated in $\BMx$ have to
be reexpressed as multipolar series in the relative variable
$\BMr^i = x^i_1 - x^i_2$, as, for example:
\begin{eqnarray}
\fl \dot{x}^k B_{k}^{(a)} (\BMx,t) +  B_{0}^{(a)} (\BMx,t)  
\nonumber\\[2mm]
\lo= \Ds \frac{m_1}{M} \left[ \dot{x}_1^k B_k^{(a)}(\BMx_1,t) 
                 + B_0^{(a)}(\BMx_1,t) \right]
      +\frac{m_2}{M} \left[ \dot{x}_2^k B^{(a)}_k(\BMx_2,t) 
                 + B^{(a)}_0(\BMx_2,t) \right] 
\nonumber \\[2mm]
+\Ds \sum_{\alpha=1}^{\infty} \frac{\mu^\alpha}{\Theta}
       \left[ 
       \left(\frac{1}{m_2}\right)^\alpha
          \left( \dot{x}_2^k B^{(a)}_{k,i_1\cdots i_\alpha}(\BMx_2,t) 
           + B^{(a)}_{0,i_1\cdots i_\alpha}(\BMx_2,t) \right)
\right. \\[2 mm] 
\Ds~~~~~~~~~~\left.
      +\left(\frac{-1}{m_1}\right)^\alpha
          \left( \dot{x}_1^k B^{(a)}_{k,i_1\cdots i_\alpha}(\BMx_1,t) 
           + B^{(a)}_{0,i_1\cdots i_\alpha}(\BMx_1,t) \right)
       \right] 
\nonumber \\[2 mm] 
\Ds~~~~~~~~~~\cdot~(x_1^{i_1}-x_2^{i_1})
             \cdots (x_1^{i_\alpha}-x_2^{i_\alpha}) 
~~~~.
\nonumber
\end{eqnarray}
It is remarkable that our constructive method based on Noether's 
theorem and Utiyama procedure leads to a {\it non-local} coupling to the 
{\it individual constituents} of the 
dynamical system even if the masses 
(sources of the Galilean gravitational
fields) are localized at their positions. 


\section{The 2-times harmonic oscillator with center-of-mass}

A description of the non-relativistic harmonic oscillator in
a 16-dimensional phase space coordinatized by $t_u$, $E_u$, $\vec{x_u}$,
$\vec{p_u}$ $(u=1,2;\;\; \{t_u,E_v\} = -\delta_{uv}$, 
$\{ x_u^i , p_{vj} \} = \delta_{uv} \delta^i_j )$ can be easily obtained by
studying the non-relativistic limit of the Todorov-Komar-Droz Vincent
relativistic harmonic oscillator. The latter is described by two
{\it first-class} constraints in the 16 variables
$x_u^\mu , p_{u\mu}$, 
and by a mutual action-at-a-distance
interaction that is {\it instantaneous in the center-of-mass frame} (see
ref\cite{LL-C}). 
The non-relativistic counterpart of this system is described by the
following two {\it first-class} constraints (the Hamiltonian is zero)
\Be
\left\{
\Ba{ll}
\Ds \bar{\chi}_a = E_a - \frac{1}{2m_a} 
                    ( \delta^{ij} p_{ai} p_{aj} 
                     +\mu k \delta_{ij} \rho^i \rho^j ) \simeq 0 
  &,~~ a=1,2 \\[2 mm]
\{ \bar{\chi}_1 , \bar{\chi}_2 \} = 0 ~~,
\Ea
\right.
\nome{6.1}
\Ee
where  $\rho^i$ is defined by 
\Be
\rho^i = r^i - \frac{t_1 - t_2}{M} \delta^{ij} P_j ~~, 
\nome{6.2}
\Ee
and $P_i=p_{1i}+p_{2i}$. \par
\vspace{2mm}

Besides eqs.\EqRef{2.2} and \EqRef{2.4}, we introduce 
the following definitions:
\Be
\fl
\left\{
\Ba{rcl}
  t &=& \Ds \frac{m_1 t_1 + m_2 t_2}{M} \\[4 mm]
 t_R&=& t_1 - t_2                       \\[4 mm]
 E  &=& E_1 + E_2                       \\[4 mm]
 E_R&=& \Ds \frac{m_2 E_1 - m_1 E_2}{M}
\Ea
\right.
\qquad
\qquad
\left\{
\Ba{rcl}
 t_1 &=& \Ds t + \frac{m_2}{M} t_R \\[4 mm]
 t_2 &=& \Ds t - \frac{m_2}{M} t_R \\[4 mm]
 E_1 &=& \Ds \frac{m_1}{M} E + E_r \\[4 mm]
 E_2 &=& \Ds \frac{m_2}{M} E - E_r 
~~~,
\Ea
\right.
\nome{6.3}
\Ee
and, for the sake of simplicity, the following 
linear combinations of the $\bar\chi_u$'s:
\Be
\fl
\left\{
\Ba{rcl}
 \bar{\chi}_{+} &=& \bar{\chi}_1 + \bar{\chi}_2
        = E - \frac{1}{2\mu} \delta^{ij} \pi_{i} \pi_{j} 
        +\frac{k}{2} \delta_{ij} \rho^i \rho^j  \simeq 0 
\\[2 mm]
 \bar{\chi}_{-} &=& \Ds { m_2 \bar{\chi}_1 - m_1 \bar{\chi}_2
                         \over M} \\[2mm]
    &=& E_r - \frac{1}{M}  \delta^{ij} P_i \pi_j
            - \frac{m_2 - m_1}{M} 
    \left( \frac{1}{2\mu} \delta^{ij} \pi_{i} \pi_{j} 
          +\frac{k}{2} \delta_{ij} \rho^i \rho^j ) 
    \right) \simeq 0 ~~.
\Ea
\right.
\nome{6.4}
\Ee
For $t_R=0$, one has $\bar{\chi}_{+} = E - \bar{H} \simeq 0$
and, correspondingly, the usual  Hamiltonian formulation
for a non-relativistic system invariant under
time reparameterization.

The Galilei generators in the Hamiltonian picture are now:
\Be
\fl \left\{
\Ba{rcl}
 E   &=& E_1 + E_2 \\
 P_i &=& p_{1i} + p_{2i} \\
 J_i &=& c_{ij}^{~~k} x_1^j p_{1k} + c_{ij}^{~~k} x_2^j p_{2k}
       = c_{ij}^{~~k} x^j P_k + c_{ij}^{~~k} r^j \pi_k \\
 K_i &=& m_1\delta_{ij} x^j_1 - t_1 p_{1i} 
        +m_2\delta_{ij} x^j_2 - t_1 p_{2i}
       = M \delta_{ij} x^j - t P_i - t_R \pi_i 
~~.
\Ea
\right.
\nome{6.5}
\Ee
Of course, these generators are constants of the motion, having zero  Poisson
brackets with both the constraints and, consequently, with
the Dirac Hamiltonian $H_D=\lambda_1 \chi_1 +\lambda_2 \chi_2$. 

 The generators of the $SU(3)\otimes U(1)$ dynamical symmetry are now:
\Be\left\{
\Ba{rcl}
 \tilde{\bar{S}}_a    &=& c_{ab}^{~~c} \rho^b \pi_c \\
 \tilde{\bar{N}}_{ab} &=& \Ds \frac{1}{2\sqrt{\mu k}} \pi_a \pi_b
                             +\frac{\sqrt{\mu k}}{2} \delta_{ac} \delta_{bd}
                                   \rho^c \rho^d ~~.
\Ea
\right.
\nome{6.6}
\Ee
They satisfy the  algebra  \EqRef{2.15} and
have zero Poisson brackets with the two constraints $\bar{\chi}_\pm$, 
because 
 $\{ \vec{\rho} , E_R - {\vec{p}\cdot\vec{\pi} \over M } \} =0$.
For $t_R =0$, they assume the expression \EqRef{2.13},\EqRef{2.14}
 of the one-time theory.
Again, we have the structure of a semi-direct product of the two
algebras \EqRef{6.5} and \EqRef{6.6}, since eqs. \EqRef{2.19} are still valid.

The generator 
$G_U = \theta^a \tilde{\bar{S}}_a + \xi^{ab} \tilde{\bar{N}}_{ab}$ 
induces the variation
\Be
\left\{
\Ba{rcl}
 \bar{\delta} x^i    &=&\Ds  \frac{t_r}{M} \delta^{ij} 
                          c_{jk}^{~~l} \theta^k \pi_l
                        -\frac{t_r}{M} \sqrt{\mu k}  
                         \xi^{ij} \delta_{jk} \rho^{k} 
\\[2 mm] 
 \bar{\delta} \rho^a &=&\Ds  c_{bc}^{~~a} \theta^b \rho^c
                        +\frac{1}{\sqrt{\mu k}} \xi^{ab} \pi_b   ~~.
\Ea
\right.
\nome{6.7}
\Ee
Now, since   $\{ x^i , \rho^j \} = - \frac{t_R}{M} \delta^{ij} \neq
0$, the {\it center-of-mass} coordinates   
invariant under the transformation of the dynamical 
symmetry are 
\Be
 \tilde{x}^i = x^i - t_R \frac{ \delta^{ij} \pi_j}{M} 
~~.
\nome{6.8}
\Ee
Then, performing the canonical transformation
\Be
\left[ \Ba{lcl} t     &,& E     \\
                t_R   &,& E_R   \\
                p_i   &,& x^i   \\
                \pi_i &,& r^i   \Ea\right] 
\longmapsto 
\left[ \Ba{lcl} 
t     &,& E  \\
t_R   &,& \tilde{E}_R = E_R - \frac{ \delta^{ij}  p_i \pi_j}{M}  \\
p_i   &,& \tilde{x}^i = x^i - t_R \frac{ \delta^{ij} p_j  }{M}   \\
\pi_i &,& \rho^i      = r^i - t_R \frac{ \delta^{ij} \pi_j}{M}   
\Ea\right] ~~,
\qquad\qquad 
\Ee
the canonical generators, associated to Galilei boosts and  rotations,
become \mbox{$\bar{K}_i = M \delta_{ij} \tilde{x}^j - t p_i$} 
and 
\mbox{$\bar{J}_i = c_{ij}^{~~k} \tilde{x}^j p_k + c_{ij}^{~~k} \rho^j\pi_k$},
while the constraint  $\bar{\chi}_{-}\simeq 0$ becomes 
\mbox{$\tilde{\bar{\chi}}_{-} = \tilde{E}_R - { m_1 -m_2 \over 2 M
\mu\simeq 0}
( \delta^{ab} \pi_a\pi_b + \mu k \delta_{ab} \rho^a \rho^b ) \simeq 0$}
and the constraint $\tilde{\bar{\chi}}_{+}\simeq 0$ is left 
unchanged.

As shown in  \cite{LL-C},  the constraints \EqRef{6.1}
follow  from the {\it singular} Lagrangian
\Be
\Ba{rcl}
\Lc^\star(\tau) &=& \Ds \left[ 1 + \frac{k t_R^2}{M^2} 
         \frac{m_1 m_2}{t_1^\prime t_2^\prime}
         \left( \frac{t_1^\prime}{m_1}+\frac{t_2^\prime}{m_2}\right)
         \right]^{-1} \cdot \\[4 mm]
    & &\Ds \cdot \left\{ \frac{m_1}{2 t_1^\prime} \delta_{ij} V_1^i V_1^j
         +\frac{m_2}{2 t_2^\prime} \delta_{ij} V_2^i V_2^j
\right.
\\[3 mm]
 &&\left. ~~
         +\frac{k t_R^2}{M^2} \frac{m_1 m_2}{t_1^\prime t_2^\prime}
         \left( \frac{t_1^\prime}{m_1}+\frac{t_2^\prime}{m_2}\right)
         \delta_{ij} (V_1^i - V_2^i) (V_1^j - V_2^j)
         \right\} \\[4 mm]
    & & \Ds - \frac{k}{2} \left( \frac{t_1^\prime}{m_1}
                                +\frac{t_2^\prime}{m_2}\right)
         \delta_{ij} \rho_i \rho_j ~~,
\Ea  
\nome{6.9}
\Ee
where $f^\prime (\tau)= \frac{d}{d\tau} f(\tau)$ and
\Be
  V_a^i = x_a^{\prime i} + \frac{k t_R}{M}
         \left(\frac{t_1^\prime}{m_1}+\frac{t_2^\prime}{m_2}\right)
~~. \nome{6.10}
\Ee
The Action $\Ac^\star = \int d\tau ~\Lc^\star(\tau)$ clearly reduces to the
Action generated by the Lagrangian \EqRef{2.1} for $t_R =0$, i.e. $t_1=t_2=t$.


\section{Gauging the $SU(3)\otimes U(1)$ group in the two-times theory}

Having the relativistic case in view (``Todorov-Komar-Droz Vincent''
harmonic oscillator \cite{Todorov,Droz}), it is instrumental 
to discuss also the {\it two-times} formulation.

Once the constraints $\bar{\chi}_\pm$ are expressed as 
$\tilde{\bar{\chi}}_\pm$, in terms of the natural canonical basis 
\( [t,E, t_R,\tilde{E}_R, \tilde{\BMx},\BMP, \BMrho,\BMpi] \) 
(see eq. (6.9)), it is apparent that they 
actually depend only on $E$, $\tilde{E}_R$
and the $SU(3)\otimes U(1)$ {\it invariant} 
$\delta^{ij} \pi_{i} \pi_{j} +k\delta_{ij} \rho^i \rho^j$.
Thus, their $SU(3)\otimes U(1)$ invariance is manifest. 
Consequently, we will now
require that $\tilde{\bar{\chi}}_\pm$ and the Dirac Hamiltonian 
be modified in such a way that they remain invariant also under the 
transformations generated by the generators 
\( G_{U} \equiv \theta^i \tilde{\bar{S}}_i + \xi^{ij}
\tilde{\bar{N}}_{ij} \), localized at 
new ``event'' $[\tilde{\BMx},t]$. Now under  
$G_{U}(\tilde{\BMx},t) \equiv  \theta^i (\tilde{\BMx},t) \tilde{\bar{S}}_i
+\xi^{ij} (\tilde{\BMx},t) \tilde{\bar{N}}_{ij}$, besides
$\bar{\delta} \BMrho \neq 0$,
$\bar{\delta} \BMpi \neq 0$,  we have also $\bar{\delta} \BMP \neq0$, 
(as in Eq.\EqRef{3.13}) and $\delta E\neq 0$,
while the variations of the $SU(3)\otimes U(1)$ 
generators maintain the form 
\EqRef{3.4} with $\BMr$ replaced by $\BMrho$. 
Then, in analogy to eqs.(3.29) and (3.30)
of Section {\bf 3}, the invariance is recovered
by adopting the minimal coupling
\Be
\Ba{rcl}
P_i &\mapsto& \Ds \Pc_i =  P_i 
                            - B_{i}^{(a)}(\tilde{\BMx},t)  S_a
                            - B_{i}^{(ab)}(\tilde{\BMx},t) N_{ab} 
\\[2 mm]
E &\mapsto& \Ds \Ec  =  E   - B_{0}^{(a)}(\tilde{\BMx},t)  S_a
                            - B_{0}^{(ab)}(\tilde{\BMx},t) N_{ab}  ~~,
\Ea
\nome{7.1}
\Ee
and the requirement that the Yang-Mills fields $B^{(a)}_k$, $B^{(a)}_0$, 
$B^{(ab)}_k$, $B^{(ab)}_0$
transform as in eqs.\EqRef{3.26} with $\vec{x}$ replaced 
by $\vec{\tilde{x}}$. 

Therefore, we get the invariant constraints 
\Be
\fl
\left\{
\Ba{rcl}
  \Ds \bar{\chi}_{+}^{\SSs SU(3)\otimes U(1)} 
        &=&\Ds E  - B_{0}^{(a)}(\tilde{\BMx},t)  S_a
               - B_{0}^{(ab)}(\tilde{\BMx},t) N_{ab}  \\
     & &\Ds - \frac{1}{2M} \delta^{ij}
            ( P_i - B_{i}^{(a)}  (\tilde{\BMx}) \tilde{\bar{S}}_a
                  - B_{i}^{(ab)} (\tilde{\BMx}) \tilde{\bar{N}}_{ab} )
\\[2 mm] 
     & &\Ds ~~~~~~~ ( P_j - B_{j}^{(a)}  (\tilde{\BMx}) \tilde{\bar{S}}_a
                    - B_{j}^{(ab)} (\tilde{\BMx}) \tilde{\bar{N}}_{ab} )

\\[2 mm]
    & & \Ds -\frac{1}{2\mu} (\delta^{ij} \pi_i \pi_j
             +\mu k\delta_{ij} \rho^i \rho^j ) \simeq 0  
\\[2 mm]
\Ds \bar{\chi}_{-}^{\SSs SU(3)\otimes U(1)} &=& \Ds  \tilde{E}_R - \frac{m_1-m_2}{2M\mu}
                     (\delta^{ij} \pi_i \pi_j
                     +\mu k\delta_{ij} \rho^i \rho^j ) \simeq 0 
\Ea
\right.
\nome{7.2}
\Ee
and we still have the integrability condition of 
the multi-temporal theory:
\Be
  \{ \bar{\chi}_{+}^{\SSs SU(3)\otimes U(1)} , 
  \bar{\chi}_{-}^{\SSs SU(3)\otimes U(1)} \} =0 ~~.
\nome{7.3}
\Ee
Were we able  to perform the inverse Legendre transformation
independently
of the specific functional form of the external {\it gauge} fields, we  
could find a putative Lagrangian $\Lc^\star_{SU(3)\otimes U(1)}$
which would be the counterpart of
the free Lagrangian $\Lc^\star$ \EqRef{6.9} with {\it non-minimally}
coupled external fields.
Yet,  this is not the case as it is clear from the fact that 
the above procedure would amount to solving the following equations
for $\BMP$, $\BMpi$, $\lambda_{+} (\tau )$,
$\lambda_{-} (\tau )$, as function of $\der{x^i}{\tau}$, $\der{t}{\tau}$,
$\der{r^a}{\tau}$, $\der{t^i_R}{\tau}$:
\Be
\fl
\left\{
\Ba{rcl}
 \Ds \der{x^i}{\tau} &=& \{{x}^i  , \bar{H}_D \} 
\\[1 mm]
       &=&\Ds  - {\lambda_{+} (\tau) \over M } \delta^{ij}
                 \big[ P_j - B_{j}^{(a)}  (\tilde{\BMx}) \tilde{\bar{S}}_a
                       - B_{j}^{(ab)} (\tilde{\BMx}) \tilde{\bar{N}}_{ab} 
                 \big] 
\\[3 mm]
       & &\Ds    +\mbox{} \lambda_{-} (\tau) 
              \left[ \delta^{ij} \frac{{\pi}_j}{M}
              + \frac{t_r}{M} \frac{m_1-m_2}{\mu M} k \rho^i
              \right]
              +\mbox{} \lambda_{+} (\tau)  
                \frac{t_r k}{\mu M}  \rho^i
\\[3 mm]
 \Ds \der{r^i}{\tau} &=& \Ds \{ {r}^i , \bar{H}_D \} 
\\[1 mm]
       &=&\Ds  - {\lambda_{+} (\tau) \over M } \delta^{ij}
                 \big[ P_i - B_{i}^{(a)}  (\tilde{\BMx}) \tilde{\bar{S}}_a
                       - B_{i}^{(ab)} (\tilde{\BMx}) \tilde{\bar{N}}_{ab} 
                 \big] 
                 \\[3 mm] 
        & & \Ds ~~~~~\cdot~
               \left[ - B_{i}^{(b)}  (\tilde{\BMx}) c_{bc}^{~~a} \rho^c
                     - B_{i}^{(ab)} (\tilde{\BMx}) \frac{1}{\sqrt{k}} \pi_b 
               \right] 
                 \\[3 mm] 
        & & \Ds - \mbox{} \lambda_{+} (\tau) 
                 \left[ - B_{0}^{(b)}  (\tilde{\BMx}) c_{bc}^{~~a} \rho^c
                     - B_{0}^{(ab)} (\tilde{\BMx}) \frac{1}{\sqrt{k}} \pi_b 
                 \right] 
                 \\[3 mm] 
        & &\Ds  - \left( {\lambda_{+} (\tau) \over \mu }  
                      + \lambda_{-} (\tau){m_2 - m_1 \over 2 M\mu }
                \right)
                \delta^{ij} \pi_j 
              +\mbox{} \lambda_{-} (\tau) 
                 \frac{\delta^{ij} P_j}{M}
\\[3 mm]
        & &\Ds    -\mbox{} \lambda_{+} (\tau) 
                      \frac{t_R \delta^{ij}}{M}
                 \left[ \partder{B_{0}^{(b)}(\tilde{\BMx})}{\tilde{x}^j} 
                        c_{bc}^{~~a} \rho^c
                     + \partder{B_{0}^{(ab)}(\tilde{\BMx})}{\tilde{x}^j} 
                       \frac{1}{\sqrt{k}} \pi_b 
                 \right] 
\\[3 mm]
        & &\Ds    -\mbox{} \lambda_{+} (\tau) 
                      \frac{t_R \delta^{ij}\delta^{kl}}{M^2}
                 \left[ \partder{B_{k}^{(b)}(\tilde{\BMx})}{\tilde{x}^j} 
                        c_{bc}^{~~a} \rho^c
                     + \partder{B_{k}^{(ab)}(\tilde{\BMx})}{\tilde{x}^j} 
                       \frac{1}{\sqrt{k}} \pi_b 
                 \right]
\\[3 mm] & &\Ds ~~~~~~~~\cdot ~ 
                 \left[ P_l - B_{l}^{(a)}  (\tilde{\BMx}) \tilde{\bar{S}}_a
                       - B_{l}^{(ab)} (\tilde{\BMx}) \tilde{\bar{N}}_{ab} 
                 \right] 
\\[3 mm]
 \Ds \der{t}{\tau}   &=& \Ds \{ t , \bar{H}_D \} 
                      = - \lambda_{+} (\tau) \\[3 mm]
 \Ds \der{t_R}{\tau} &=&\Ds  \{ t_R , \bar{H}_D  \} 
                      = - \lambda_{-} (\tau) \\[2 mm]
 \Ds \bar{\chi}_{+}^{\SSs SU(3)\otimes U(1)} &=& 0 \\[2 mm]
 \Ds \bar{\chi}_{-}^{\SSs SU(3)\otimes U(1)} &=& 0 ~~,
\Ea
\right.
\nome{7.4}
\Ee
(where $\bar{H}_D = \lambda_{+} \bar{\chi}_{+}^{\SSs SU(3)\otimes U(1)}
                  +\lambda_{-} \bar{\chi}_{-}^{\SSs SU(3)\otimes U(1)}$
is the Dirac's Hamiltonian).
In particular, one would have to solve for, e.g., the variable
$\BMpi$, which is itself an argument
of the {\it gauge} fields through $\tilde{\BMx}$.

On the other hand, a Lagrangian description of the {\it
gauge}-invariant  system can be obtained in a natural and simpler 
way by fully resorting to {\it new} position {\it and} 
energy particle variables 
$[ \tilde{x}_1^i, \tilde{x}_2^i, \tilde{E}_1 , \tilde{E}_2 ]$
defined by
\Be
\left\{
\Ba{rcl}
 t_1           &=&\Ds   t + \frac{m_2}{M} t_R 
\\[2 mm] 
 \tilde{E}_1   &=&\Ds   E_1 - \frac{\delta^{ij}p_1 \pi_j}{M}   
\\[2 mm] 
 t_2           &=&\Ds   t - \frac{m_2}{M} t_R
\\[2 mm] 
 \tilde{E}_2   &=&\Ds   E_2 + \frac{\delta^{ij}p_1 \pi_j}{M}
\\[2 mm] 
 \tilde{x}_1^i &=&\Ds   \tilde{x}^i  + \frac{m_2}{M} \rho^i
\\[2 mm] 
 p_{1i}        &=&\Ds  \frac{m_1}{M} P_i + \pi_i 
\\[2 mm] 
 \tilde{x}_2^i &=&\Ds   \tilde{x}^i  - \frac{m_1}{M} \rho^i
\\[2 mm] 
 p_{2i}        &=&\Ds  \frac{m_2}{M} P_i - \pi_i 
\Ea
\right.
\qquad
\left\{
\Ba{rcl}
 t           &=&\Ds   {m_1 t_1 + m_2 t_2 \over M}
\\[2 mm] 
 E           &=&\Ds   \tilde{E}_1+\tilde{E}_2 
\\[2 mm] 
 t_R         &=&\Ds   t_1 - t_2 
\\[2 mm] 
 \tilde{E}_R &=&\Ds {m_2 \tilde{E}_1 - m_1 \tilde{E}_2 \over M} 
\\[2 mm] 
 \tilde{x}^i &=&\Ds {m_1 \tilde{x}_1^i + m_2 \tilde{x}_2^i \over M} 
\\[2 mm] 
 p_{i}       &=&\Ds p_{1i} + p_{2i}
\\[2 mm] 
 \rho^i      &=&\Ds \tilde{x}^i_1 - \tilde{x}^i_2 
\\[2 mm] 
 \pi_i       &=&\Ds {m_2 p_{1i} - m_1 p_{2i} \over M}  
~~.
\Ea
\right.
\nome{7.5}
\Ee
First of all, this is easily done for the uncoupled system. Actually,
the original constraints (\ref{6.1}) can be rewritten as:
\Be
\fl \left\{
\Ba{rcl}
 \bar{\chi}_{+} &=& \tilde{\bar{\chi}}_1 + \tilde{\bar{\chi}}_2
                 =\Ds E 
                     -\frac{1}{2M} \delta^{ij} P_{i} P_{j} 
                     - \frac{1}{2\mu} 
                    ( \delta^{ij} \pi_{i} \pi_{j} 
                    +k\delta_{ij} \rho^i \rho^j ) \simeq 0 \\[2 mm]
 \bar{\chi}_{-} &=& \Ds { m_2 \tilde{\bar{\chi}}_1 
                        - m_1 \tilde{\bar{\chi}}_2
                         \over M}
                 = \tilde{E}_R 
                   - \frac{m_2 - m_1}{2\mu M} 
                    ( \delta^{ij} \pi_{i} \pi_{j} 
                    +k\delta_{ij} \rho^i \rho^j ) \simeq 0 ~~,
\Ea
\right.
\nome{7.6}
\Ee
or
\Be
\fl \left\{
\Ba{ll}
\Ds \tilde{\bar{\chi}}_u = \tilde{E}_u - \frac{1}{2m_u} 
           \left( \delta^{ij} p_{ui} p_{uj} 
                 +\mu k \delta_{ij} \rho^i \rho^j ) 
           \right) \simeq 0 
  &u=1,2 \\[2 mm]
\{ \bar{\chi}_1 , \bar{\chi}_2 \} = 0 ~~.
\Ea
\right.
\nome{7.7}
\Ee
Performing the inverse Legendre transformation,
one obtains a Lagrangian dependent on
$[ t_u(\tau), t_u^\prime(\tau), 
\tilde{\BMx}^i_u(\tau), \tilde{\BMx}_u^{\prime i}(\tau) ]$
or, what is the same,
$[t_u(\tau), t_a^\prime(\tau), \tilde{\BMx}^i(\tau), 
\tilde{\BMx}^{\prime i}(\tau), \BMrho(\tau), \BMrho^\prime(\tau) ]$:
\Be
\fl\Ba{rcl}
  \Lc^{\star} &=& \Ds \frac{M}{2} 
                \frac{\delta_{ij} \mbox{$\primato{\tilde{x}}$}^i 
                                  \mbox{$\primato{\tilde{x}}$}^j
                     }{\primato{t}}
    +  \frac{\mu}{2} 
                \frac{\delta_{ab} \primato{\rho}^a 
                                  \primato{\rho}^b
                     }{\primato{t}+\frac{m_2-m_1}{M} \primato{T}_R}
     -\mbox{} \frac{k}{2}
           \left({\primato{t}+\frac{m_2-m_1}{M} \primato{T}_R}
           \right) \delta_{ab} \rho^a \rho^b  ~~.
\Ea
\nome{7.8}
\Ee
On the other hand, the constraints of the coupled system
(given by eq. (7.2)) determine the equations of motion
\Be
\left\{
\Ba{rcl}
 \mbox{$\primato{\tilde{x}}$}^i &=& \{ \tilde{x}^i , \bar{H}_D \} 
\\
            &=& - {\lambda_{+} (\tau) \over M } \delta^{ij}
                 ( P_j - B_{j}^{(a)}  (\tilde{\BMx}) \tilde{\bar{S}}_a
                       - B_{j}^{(ab)} (\tilde{\BMx}) \tilde{\bar{N}}_{ab} ) 
                 \\[3 mm]
 \primato{\rho}^a &=& \{ \rho^a , \bar{H}_D  \}
\\
            &=& - {\lambda_{+} (\tau) \over M } \delta^{ij}
                 ( P_i - B_{i}^{(a)}  (\tilde{\BMx}) \tilde{\bar{S}}_a
                       - B_{i}^{(ab)} (\tilde{\BMx}) \tilde{\bar{N}}_{ab} ) 
                 \\[3 mm] 
            & & ~~~( - B_{i}^{(b)}  (\tilde{\BMx}) c_{bc}^{~~a} \rho^c
                     - B_{i}^{(ab)} (\tilde{\BMx}) \frac{1}{\sqrt{\mu k}} \pi_b ) 
                 \\[3 mm] 
            & & - \mbox{} \lambda_{+} (\tau) 
                   ( - B_{0}^{(b)}  (\tilde{\BMx}) c_{bc}^{~~a} \rho^c
                     - B_{0}^{(ab)} (\tilde{\BMx}) \frac{1}{\sqrt{\mu k}} \pi_b ) 
                 \\[3 mm] 
            & & - \left( {\lambda_{+} (\tau) \over \mu }  
                      + \lambda_{-} (\tau){m_2 - m_1 \over 2 M\mu }
                \right)
                \delta^{ij} \pi_j \\[3 mm]
 \primato{t}  &=& \{ t , \bar{H}_D \} = - \lambda_{+} \\[3 mm]
 \primato{t}_R &=& \{ t_R , \bar{H}_D  \} = - \lambda_{-} (\tau) \\[1 mm]
 \bar{\chi}_{+}^{\SSs SU(3)\otimes U(1)} &=& 0 \\[1 mm]
 \bar{\chi}_{-}^{\SSs SU(3)\otimes U(1)} &=& 0 ~~,
\Ea
\right.
\nome{7.9}
\Ee
which can  be explicitly solved for 
$\BMP$, $\BMpi$, $\lambda_{+}(\tau)$, $\lambda_{-}(\tau)$
as functions of $[t_u(\tau), t_u^\prime(\tau), 
\tilde{\BMx}^i_u(\tau), \tilde{\BMx}_u^{\prime i}(\tau)]$

Thus, the inverse Legendre transformation can be carried out. 
As a matter of fact, by defining, in analogy to what done in section 
{\bf3}, the quantities 
\Bea
\fl 
D \rho^a &\equiv& \rho^{\prime a} + c_{bc}^{~~a}     
     \left( {\tilde{x}}^{\prime k} B_k^{(b)} (\tilde{\BMx},t) 
           +\primato{t}       B_0^{(b)} (\tilde{\BMx},t) 
     \right) \rho^c
\nome{7.10}  
\nome{7.11}
\\[2 mm]
\fl
\tilde{\Xi}^{ab} (\tilde{\BMx},t,t_R) &\equiv&
    \delta^{ab} + 
      \frac{\mu}{\sqrt{\mu k}} 
      \frac{1}{\primato{t}+\frac{m_2-m_1}{M}\primato{t}_R}
    \left( {\tilde{x}}^{\prime k} B_k^{(ab)} (\tilde{\BMx},t) 
          +\primato{t}       B_0^{(ab)} (\tilde{\BMx},t) 
    \right)    ~~,
\nonumber
\Eea
so that
\Be
\tilde{\Xi}^{ab} (\tilde{\BMx},t,t_R) \pi_{j}
   =\Ds  \mu
      \frac{1}{\primato{t}+\frac{m_2-m_1}{M}\primato{t}_R}
     D \rho^i   ~~,
\Ee
it follows
\Be
\left\{
\Ba{rcl}
 \lambda_{+} &=& - \primato{t} \\
 \lambda_{-} &=& - \primato{t}_R \\
 \pi_a   &=& \Ds  \mu
      \frac{1}{\primato{t}+\frac{m_2-m_1}{M}\primato{t}_R}
        |\Xi^{-1} (\tilde{\BMx},t,t_R)|_{ab} ~D \rho^b  
\\[2 mm]
  P_i  &=& M \delta_{ij} \tilde{x}^{\prime j}  
          + B_{i}^{(a)}  (\tilde{\BMx}) \tilde{\bar{S}}_a
          + B_{i}^{(ab)} (\tilde{\BMx}) \tilde{\bar{N}}_{ab} ~~,
\Ea
\right.
\nome{7.12}
\Ee
In conclusion, the Lagrangian of the coupled system, 
$\tilde{\Lc}^{\star}_{SU(3)\otimes U(1)}$,
is given by:
\Be
\fl
\Ba{rcl}
  \tilde{\Lc}^{\star}_{SU(3)\otimes U(1)} 
       &=& \Ds \frac{M}{2} 
                \frac{\delta_{ij} \mbox{$\primato{\tilde{x}}$}^i 
                                  \mbox{$\primato{\tilde{x}}$}^j
                     }{\primato{t}}
             +  \frac{\mu}{2} 
                \frac{|\Xi^{-1} (\tilde{\BMx},t,t_R)|_{ab} 
                      D{\rho}^a ~D{\rho}^b
                     }{\primato{t}+\frac{m_2-m_1}{M} \primato{T}_R}
\\[4 mm]
      & &\Ds  ~-\mbox{} \frac{k}{\mu}
           \left({\primato{t}+\frac{m_2-m_1}{M} \primato{t}_R}
           \right) |\Xi(\tilde{\BMx},t,t_R)|^{ab} 
                    \delta_{ac} \delta_{bd} \rho^c \rho^d ~~~.
\Ea
\nome{7.13}
\Ee

  The main drawback of the Lagrangian 
$\tilde{\Lc}^{\star}_{SU(3)\otimes U(1)}$ 
is that it cannot be transformed in practice into a putative {\it
non-minimally} coupled Lagrangian  ${\Lc}^{\star}_{SU(3)\otimes U(1)}$,
expressed in terms of the original configurational variables 
$[\BMx_1,t_1;\BMx_2,t_2]$. Indeed, we would have to perform a 
velocity-dependent change of coordinates,
which would be the {\it pull-back}
of the canonical transformation \EqRef{7.5}.


\section{Gauging the maximal symmetry group in the two-time theory}

In terms of  the canonical basis 
$[t_a(\tau), t_a^\prime(\tau), \tilde{\BMx}^i(\tau),
  \tilde{\BMx}^{\prime i}(\tau),\BMrho(\tau), \BMrho^\prime(\tau) ]$,
the generator of the {\it gauge} transformations of the maximal
symmetry group localized at the effective center-of-mass
$\tilde{\BMx}$
is given by:
\Be
\fl
\Ba{rcl}
 \tilde{\bar{G}}_{\Sss \Hc} = \epsilon(t) E
            + \varepsilon^i(\tilde{\BMx},t) P_i
            + \omega^i(\tilde{\BMx},t) \bar{J}_i 
            + v^i(\tilde{\BMx},t) \bar{K}_i 
            + \theta^a(\tilde{\BMx},t) \bar{S}_a 
            + \xi^{ab}(\tilde{\BMx},t) \bar{N}_{ab} ~~.
\Ea
\nome{8.1}
\Ee
Defining, as in Section {\bf 4}, 
$\tilde{\theta}^i \equiv \theta^i + \omega^i $
and $\tilde{\eta}^i \equiv \varepsilon^i(\tilde{\BMx},t) 
+ c_{jk}^{~~i} \omega^j(\tilde{\BMx},t) \tilde{x}^j 
- t  v^i(\tilde{\BMx},t)$, the transformation properties of the
configurational variables are 
\Be
\left\{
\Ba{rcl}
 \delta \tilde{x}^i &=& \Ds \tilde{\eta}^i(\tilde{\BMx},t)  
\\[2 mm]
 \delta r^a &=&  c_{bc}^{~~a} \tilde{\theta}^b(\tilde{\BMx},t) r^c
               + {\sqrt{\mu k}} \xi^{ab}(\tilde{\BMx},t) 
                 \pi_b
\\[2 mm]
 \delta t     &=& - \varepsilon(t) \\
 \delta t_R   &=& 0 ~~~.
\Ea
\right.
\nome{8.2}
\Ee
Clearly, with respect to the transformations generated by
$\tilde{\bar{G}}_{\Sss \Hc}$ we have:
\Be
{\delta}^{\Hc} ( \delta^{ij} \pi_i \pi_j 
                     + k \delta_{ij} \rho_i \rho_j ) = 0 ~~,
\nome{8.4}
\Ee
and, since we have also $\bar{\delta}^\prime \tilde{E}_R =0$, 
we obtain 
\Be
\bar{\delta}^\prime \tilde{\bar{\chi}}_{-} =
 \bar{\delta}^\prime \left[ \tilde{E}_R
 - \frac{m_1-m_2}{2M\mu} ( \delta^{ij} \pi_i \pi_j 
                          +k \delta_{ij} \rho_i \rho_j )\right]=0 ~~.
\Ee
We conclude that, because of its invariance, 
this constraint does not couple to external fields.
As to the constraints $\tilde{\bar{\chi}}_{+}$,
the only quantities with non vanishing variation are 
\Be
\Ba{rcl}
  \delta^{\Hc} P_i &=& \Ds 
            - \partder{\tilde{\eta}^{j}}{\tilde{x}^i} P_j
            - \partder{\tilde{\theta}^{a}}{\tilde{x}^i} \bar{S}_{a}
            - \partder{\xi^{ab}}{\tilde{x}^i} \bar{N}_{ab}
            - M \partder{}{\tilde{x}^i} \left[ g_{ij}v^i\tilde{x}^j\right]
\\[2 mm]
  \delta^{\Hc} E &=& \Ds 
            - \partder{\tilde{\eta}^{j}}{t} P_j
            - \partder{\tilde{\theta}^{a}}{t} \bar{S}_{a}
            - \partder{\xi^{ab}}{t} \bar{N}_{ab}
            - M \partder{}{t} \left[ g_{ij}v^i\tilde{x}^j\right]
~~~.
\Ea
\nome{8.5}
\Ee
Thus, the invariance of $\tilde{\bar{\chi}}_{+}$ under the
transformations (\ref{8.2}) can be recovered by introducing
the minimal coupling
\Be
\fl
\left\{
\Ba{rcl}
P_i &\mapsto& \Ds \Pc_a = {\bf H}_{a}^i({\tilde{\BMx}},t) 
               \left[ P_i - \frac{M}{\Theta} A_i({\tilde{\BMx}},t)
                            - B_{i}^{(b)}({\tilde{\BMx}},t)  S_b
                            - B_{i}^{(cd)}({\tilde{\BMx}},t) N_{cd}  
               \right]
\\[2 mm]
E &\mapsto& \Ds \Ec  = \frac{1}{\Theta} 
                \left[ E - \frac{M}{\Theta} A_0({\tilde{\BMx}},t)
                            - B_{0}^{(a)}({\tilde{\BMx}},t)  S_a
                            - B_{0}^{(ab)}({\tilde{\BMx}},t) N_{ab}  
                \right]
~~~,
\Ea
\right.
\nome{8.6}
\Ee
and assuming transformation rules of the {\it gauge} fields so that 
the condition 
\Be
   \delta^g \tilde{\bar{\chi}}^G_{+} 
  = \delta^g \left[ \Ec - \frac{1}{2M} \delta^{ab} \Pc_a\Pc_b
     - \frac{1}{2\mu} \left( \delta^{ab}\pi_a \pi_b 
                           + k \delta_{ab} \rho^a \rho^b
                      \right) 
    \right] = 0
\nome{8.7}
\Ee
be satisfied. \newline
\noindent The consequent transformation rule results:
\begin{eqnarray}
\hat{\delta}_0^g \Theta (t)  &=& \dot{\epsilon}(t) \Theta (t)  
\nonumber \\[2 mm]
\hat{\delta}_0^g  g_{ij} (\tilde{\BMx},t) &=& \Ds 
           - \partder{\tilde\eta^k (\tilde{\BMx},t) }{\tilde{x}^i} g_{kj}
           - \partder{\tilde\eta^k (\tilde{\BMx},t) }{\tilde{x}^i} g_{kj} 
\nonumber \\[2 mm]
\hat{\delta}_0^g A_0 (\tilde{\BMx},t)     &=& \Ds 2 \dot{\varepsilon} A_0
           - A_i \partder{\tilde\eta^i}{t}
           - \Theta \partder{}{t}
           \left[{g_{ij} v^i \tilde{x}^j}\right] 
\nonumber \\[2 mm]
\hat{\delta}_0^g A_i (\tilde{\BMx},t)     &=& \Ds \dot{\varepsilon} A_i
           - A_j \partder{\tilde\eta^j}{\tilde{x}^i}
           - g_{ij}  \partder{\tilde\eta^j}{t}
           - \Theta \partder{}{\tilde{x}^i}
           \left[{g_{jk} v^j \tilde{x}^k}\right]  ~~.
\nonumber \\[2 mm]
\hat{\delta}_0^g B_{k}^{(a)}(\tilde{\BMx},t)
             &=& \Ds   c_{bc}^{~~a} \tilde{\theta}^b (\tilde{\BMx},t) 
                       B_{k}^{(c)}(\tilde{\BMx},t)   
\nonumber \\[2 mm]
             & & \Ds  + \xi^{cd} (\tilde{\BMx},t)
                    [c_{ce}^{~~a} \delta_{df} B_{k}^{(ef)}(\tilde{\BMx},t)
                    +c_{de}^{~~a} \delta_{cf} B_{k}^{(ef)}(\tilde{\BMx},t)  ]  
\nonumber \\[2 mm]
             & & \Ds
           - \partder{\tilde\eta^i}{\tilde{x}^k} B_{i}^{(a)}(\tilde{\BMx},t) 
           - \partder{\tilde{\theta}^a (\tilde{\BMx},t)}{\tilde{x}^k}  
\nonumber \\[3 mm]
\hat{\delta}_0^g B_{0}^{(a)}(\tilde{\BMx},t)
             &=& \Ds   c_{bc}^{~~a} \tilde{\theta}^b (\tilde{\BMx},t) 
                       B_{0}^{(c)}(\tilde{\BMx},t)
\nonumber \\[2 mm]
             & & \Ds  + \xi^{cd} (\tilde{\BMx},t)
                    [c_{ce}^{~~a} \delta_{df} B_{0}^{(ef)}(\tilde{\BMx},t)
                    +c_{de}^{~~a} \delta_{cf} B_{0}^{(ef)}(\tilde{\BMx},t)  ]  
\nonumber \\[2 mm]
             & & \Ds 
           + \dot{\epsilon} B_{0}^{(a)}(\tilde{\BMx},t) 
           - \partder{\tilde\eta^i}{t} B_{i}^{(a)}(\tilde{\BMx},t) 
           - \partder{\tilde{\theta}^a (\tilde{\BMx},t)}{t}  
\nonumber \\[3 mm]
\hat{\delta}_0^g B_{k}^{(ab)}(\tilde{\BMx},t)
           &=&\tilde{\theta}^c (\tilde{\BMx},t) 
                 [ c_{cd}^{~~a} B_{k}^{(db)} (\tilde{\BMx},t)
                  +c_{cd}^{~~b} B_{k}^{(da)} (\tilde{\BMx},t) ]  
\nonumber \\[2 mm]
           & &\Ds   + \frac{1}{4} \xi^{cd}(\tilde{\BMx},t)
                 [c_{ec}^{~~a} \delta_d^b
                 +c_{ec}^{~~b} \delta_d^a
                 +c_{ed}^{~~a} \delta_c^b
                 +c_{ed}^{~~b} \delta_c^a] B_{k}^{(e)}(\tilde{\BMx},t) 
\nonumber \\[2 mm]
           & &\Ds
           - \partder{\tilde\eta^i}{\tilde{x}^k} B_{i}^{(ab)}(\tilde{\BMx},t) 
           - \partder{\xi^{ab}(\tilde{\BMx},t)}{\tilde{x}^k}
\nonumber \\[3 mm]
\hat{\delta}_0^g B_{0}^{(ab)}(\tilde{\BMx},t)
           &=&\tilde{\theta}^c (\tilde{\BMx},t) 
                  [ c_{cd}^{~~a} B_{0}^{(db)} (\tilde{\BMx},t)
                   +c_{cd}^{~~b} B_{0}^{(da)} (\tilde{\BMx},t) ]  
\nonumber \\[2 mm]
           & &\Ds   + \frac{1}{4} \xi^{cd}(\tilde{\BMx},t)
                 [c_{ec}^{~~a} \delta_d^b
                 +c_{ec}^{~~b} \delta_d^a
                 +c_{ed}^{~~a} \delta_c^b
                 +c_{ed}^{~~b} \delta_c^a] B_{0}^{(e)}(\tilde{\BMx},t) 
\nonumber \\[2 mm]
           & &\Ds
           + \dot{\epsilon} B_{0}^{(ab)}(\tilde{\BMx},t) 
           - \partder{\tilde\eta^i}{t} B_{i}^{(ab)}(\tilde{\BMx},t) 
           - \partder{\xi^{ab}(\tilde{\BMx},t)}{t}
~~~.
\nonumber 
\end{eqnarray}
Moreover, it can be checked that  the integrability condition of
the multi-temporal formalism, $\{ \chi^{\Hc}_{+} , \chi^{\Hc}_{-} \} =0$,
still holds.

Using the definitions \EqRef{7.11} of Section {\bf 7}, in which the
time reparametrization field $\Theta(t)$ must now be introduced, i.e.:
\Bea
\fl
D \rho^a &=& \rho^{\prime a} + c_{bc}^{~~a}     
     \left( {\tilde{x}}^{\prime k} B_k^{(b)} (\tilde{\BMx},t) 
           +\primato{t}       B_0^{(b)} (\tilde{\BMx},t) 
     \right) \rho^c
\\[2 mm]
\fl
\tilde{\Xi}^{ab} (\tilde{\BMx},t,t_R) &=&
    \delta^{ab} + 
      \frac{\mu}{\sqrt{\mu k}} 
      \frac{1}{\Theta(t)\primato{t}+\frac{m_2-m_1}{M}\primato{t}_R}
    \left( {\tilde{x}}^{\prime k} B_k^{(ab)} (\tilde{\BMx},t) 
          +\primato{t}       B_0^{(ab)} (\tilde{\BMx},t) 
    \right)    ~~,
\nonumber
\Eea
we get the final result, as a functional of the variables
$[\tilde{\BMx},\BMrho,t,t_R]$, in the form:
\Be
\fl
\Ba{rcl}
  \tilde{\Lc}^{\star}_{\Hc} &=& \Ds \frac{M}{2} 
                \frac{g_{ij}(\tilde{\BMx},t)  \mbox{$\primato{\tilde{x}}$}^i 
                                  \mbox{$\primato{\tilde{x}}$}^j
                     }{\Theta(t) \primato{t}}
              + M~A_i(\tilde{\BMx},t)  \mbox{$\primato{\tilde{x}}$}^i
              + M~\Theta(t) ~A_0(\tilde{\BMx},t) \primato{t}
\\[4 mm]
      & &\Ds ~+  \frac{\mu}{2} 
                \frac{|\Xi^{-1} (\tilde{\BMx},t,t_R)|_{ab} 
                      D{\rho}^a ~D{\rho}^b
                     }{\Theta(t)\primato{t}+\frac{m_2-m_1}{M} \primato{T}_R}
\\[4 mm]
      & &\Ds ~-\mbox{} \frac{k}{\mu}
           \left({\Theta(t)\primato{t}+\frac{m_2-m_1}{M} \primato{t}_R}
           \right) |\Xi(\tilde{\BMx},t,t_R)|^{ab} 
                    \delta_{ac} \delta_{bd} \rho^c \rho^d ~~.
\Ea
\Ee
Clearly, this new Lagrangian reduces to the Lagrangian $\Lc_{\Hc}$
(\ref{5.3}) when $t=\tau$, $t_R=0$. The dynamical symmetry
transformations,  under which this Lagrangian is invariant 
({\it quasi-invariant}) are given in configuration space by:
\Be
\fl
\left\{
\Ba{rcl}
 \delta^{\Hc} \tilde{x}^i &=& \Ds \tilde{\eta}^i(\tilde{\BMx},t)  
\\[3 mm]
 \delta^{\Hc} \rho^a &=&\Ds  \epsilon(t) \dot{r}^a 
                  + c_{bc}^{~~a} \tilde{\theta}^b(\tilde{\BMx},t) \rho^c
                  + \frac{\mu}{\sqrt{\mu k}} \xi^{ab}(\tilde{\BMx},t) 
                    \frac{|\Xi^{-1}|_{bc} (\tilde{\BMx},t,t_R) D{\rho}^c 
                         }{\Theta(t)\primato{t}
                           +\frac{m_2-m_1}{M} \primato{t}_R}  ~~.
\Ea
\right.
\Ee


\section{Conclusions}

The main result of this paper is a well-defined algorithm
for characterizing the coupling of {\it extended, bounded,
non-relativistic systems} (whose binding is given by long
range forces or confining interactions), having a given
maximal dynamical symmetry group ({\it semi-direct} product
of the Galilei group times the internal maximal dynamical
symmetry group), to {\it external} gravitational and Yang-Mills
fields. This algorithm, based on a generalization of
Utiyama's procedure, has been applied to 
the non-relativistic oscillator with
{\it center-of-mass}. The explicit form of the coupling in
this system is given by the Lagrangian (5.4). Its expression
in terms of the coordinates of the individual constituents
(see eqs. 5.9,5.10) shows the non-local nature of the coupling 
to the latter,
since the algebraic approch privileges a local coupling to
the {\it center-of-mass} and {\it relative} variables. We cannot exclude
the existence of a local coupling to the constituent
particles (like the standard coupling to the free test
particles of a dust) but certain it cannot be obtained by
exploiting {\it algebraic methods} {\it \`a la} Utiyama.

It seems reasonable that the method we are proposing is
sensible for bound systems but not for a dust of free
constituents (whose maximal symmetry group is the semidirect
product of the Galilei group times $U(1)$, times
$E(3)_1,\ldots,E(3)_{N-1}$, for $N$ particles) because, in
this case, one would get a non standard coupling to the
{\it center-of-mass}, contrary to the conventional wisdom about
the treatment of a dust in general relativity or Galilean
gravity.

Our approch allows the treatment of a dust of bounded systems
by exploiting the algebraic approch for the latter and the
standard approch for a dust of {\it center-of-masses}.

We have discused also the two-time reformulation of the
Galilean harmonic oscillator in which the system is
described only by two {\it first-class} constraints 
(the canonical Hamiltonian vanishes) as a preliminary step
towards the {\it relativistic} case in which only this kind of Hamiltonian
description is available \cite{Todorov,Droz}.
In the relativistic case, in which the Poincar\'e kinematical 
group is present by definition - both in the 
case of a bilocal system\footnote{The relativistic oscillator still
possess a $SU(3)\otimes U(1)$ algebra of constants of motion.}
and for continuous systems as the Nambu-Goto
string \footnote{See the infinite-dimensional algebra of constants
of motion of reference \cite{NambuGoto}.} - the resulting 
{\it ``maximal symmetry algebra''},\ including
both kinematical and dynamical symmetries, is not a Lie algebra
but some kind of generalized W-algebra with structure constants depending
on the total four-momentum.


\ack

Roberto De Pietri wishes to thank C. Rovelli and E.T. Newman for
the hospitality kindly offered to him at the Department of Physics 
and Astronomy. Massimo Pauri would like to express his deep 
appreciation and thanks to
the {\it Center for Philosophy of Science}, for the warm and 
stimulating intellectual
atmosphere experienced there and the generous partial support obtained
during the completion
of the present work at the University of Pittsburgh.


\appendix
\section{Canonical realizations of the $U(3)=SU(3)\otimes U(1)$ group}


In standard notations, the canonical generators of 
the $U(3)=SU(3)\otimes U(1)$ group can be written as:
\begin{equation}
   X_{a~-b} =  N_{ab} + {i \over 2} S_{ab} ~~,
\nome{A.1}
\end{equation}                                     
(where $N_{ab}$, $S_{ab}$ are symmetric and 
antisymmetric real quantities, respectively) 
which satisfy the (Poisson brackets)
commutation rules:
\begin{equation}
  \{  X_{a~-b} , X_{c~-d} \} = \dot{\imath}
        ( \delta_{bc}  X_{a~-d} - \delta_{ad}  X_{c~-b} )
~~.
\nome{A.2}
\end{equation}
Expressing  $N_{ab}$ in terms of its {\it trace} and 
{\it traceless} parts
($N,~\tilde{N}_{ab}$) as: $N_{ab} = \tilde{N}_{ab} + \frac{1}{3} 
\delta_{ab} N$, the commutation rules can be rewritten 
in terms of $S_a$, $\tilde{N}_{ab}$ and $N$ as:
\begin{eqnarray}
   \{ N   , {S}_b \} &=&  0    \nonumber\\
   \{ N   , \tilde{N}_{cd} \} &=& 0 \nonumber\\
   \{ {S}_a   , {S}_b \} &=&  \epsilon_{abc} {S}_c    
\nome{A.3}\\
   \{ {S}_a   , \tilde{N}_{cd} \} &=& \epsilon_{acb} \tilde{N}_{bd}
                                     +\epsilon_{adb} \tilde{N}_{cb} 
\nonumber \\
   \{ \tilde{N}_{ab}, \tilde{N}_{cd} \} &=& \frac{1}{4} [
           \delta_{bd} \epsilon_{ace}+ \delta_{bc} \epsilon_{ade}
         + \delta_{ad} \epsilon_{bce}+ \delta_{ac} \epsilon_{bde}] {S}_e
\nonumber ~~,
\end{eqnarray}
where the $SU(3)\otimes U(1)$ group structure is explicitly
evidentiated.

According to the general theory \cite{Pauri}, a canonical
realization of a Lie group $G_r$ (of order $r$) can be
characterized in  terms of two basic schemes: 
A) the  {\it scheme A} which depends entirely on the
structure of the Lie algebra ${\cal L}_{G_r}$ 
(including its {\it cohomology}) and amounts to
a  {\it pseudo-canonization} of the  generators, in terms of
$k$ {\it invariants} and $h= (r-k)/2$ pairs of canonical
variables ({\it kernel} of the {\it scheme}); 
B) the {\it scheme B} (or {\it typical form}) which is an
array of $2n$ canonical  variables $P_i$,$Q_i$, defined by
means of a {\it canonical completion} of  the {\it scheme
A}. The {\it scheme B} allows to {\it analyze} any {\it given} 
canonical realization of $G_r$ and to {\it construct} the most 
general canonical realization of $G_r$. In what follows
we shall be interested in the {\it analysis} of {\it given} realizations
in which the explicit expressions of the generators is already
given in terms of the {\it physical} variables $p_i$,
$q_i$ (see later). 

The {\it scheme A} for the group $SU(3)\otimes U(1)$ has 
the generic form ($k=3$, $h=3$):

\vspace{0.5cm}
\Be\mbox{
\begin{tabular}{||ccc|cc|c||} 
\hline \hline
    ${\cal P}_1 $ & ${\cal P}_2$ & ${\cal P}_3$            &
    ${\Jc}_1 $ & ${\Jc}_2$ & ${\Jc}_3$           \\ 
    ${\cal Q}_1 $ & ${\cal Q}_2$ & ${\cal Q}_3$             & 
       &    &      \\
\hline
\hline
\end{tabular}
}
\nome{genU3}
\Ee
\vspace{0.5cm}

\noindent
where variables belonging to the same vertical pair are
canonical conjugated, and variables belonging to
different vertical lines commute. 
The quantities ${\Jc}_1$, ${\Jc}_2$ and ${\Jc}_3$ clearly
commute with all of the generators and are the {\it invariants} 
of the group.
Of course, any set of $k$ functional independent 
functions $\Jc_1'(\Jc_1,\cdots,\Jc_k),\cdots,
\Jc_k'(\Jc_1,\cdots,\Jc_k)$ of the {\it invariants} 
are good {\it invariants} as well.
The explicit form of the variables (\ref{genU3}), as functions of the 
generators, can be chosen as follows:
\begin{equation}
  \begin{array}{rclrcl}
{\cal P}_1 &=& S_z & 
{\cal P}_2 &=& \sqrt{\Ts\vec S^2}  \\[2 mm]
{\cal Q}_1 &=& \Ds \arctan \frac{\Ts S_y}{\Ts S_x} ~~~~~~ &  
{\cal Q}_2 &=& \Ds \arctan 
               \frac{\Ts S (\vec{S}\wedge\vec{W})_z}{
                     \Ts (\vec{S}\cdot\vec{W})~ S_z - S^2 W_z} 
~~~~~\, ,
\end{array}
\nome{A.4}
\end{equation}
where  $W_a = \tilde{N}_{ab} S^b$, while ${\cal P}_3$ and 
${\cal Q}_3$ are  (elliptic) rotational scalar functions of
their arguments:
\begin{eqnarray}
  {\cal P}_3 &=& {\cal P}_3
 ~[\vec{W}\cdot\vec{S},\vec{W}\cdot\vec{W},
   {\Jc}_1,{\Jc}_2,{\Jc}_3]
\nome{A.5} \\
  {\cal Q}_3 &=& {\cal Q}_3
 ~[\vec{W}\cdot\vec{S},\vec{W}\cdot\vec{W},
   {\Jc}_1,{\Jc}_2,{\Jc}_3]  \, ,
\nonumber
\end{eqnarray}
whose explicit expression does not matter here (see ref.
\cite{Futuro}).  

A possible choice of three independent functions of the 
{\it invariants} is:
\begin{eqnarray}
  {\Jc}_1 &=& {1 \over 2} \sum_{a=1}^3 S_a^2 
                +  \sum_{a,b=1}^3 \tilde{N}_{ab} \tilde{N}_{ba}   
 \nonumber\\
  {\Jc}_2 &=& \sum_{a,b,c=1}^3  
              \tilde{N}_{ab} \tilde{N}_{bc} \tilde{N}_{ca}
          + {3\over 4} \sum_{a,b=1}^3 \tilde{N}_{ab} S^a S^b 
\nome{A.6}\\[2 mm]
  {\Jc}_3   &=& N
~~.   \nonumber
\end{eqnarray}

A {\it scheme A} is called {\it singular} \cite{Pauri} if, 
due to some functional relations, already existing or imposed,
among the {\it invariants}, some canonical pairs  
become singular functions of the
generators and must be omitted from {\it scheme A} itself
which, then, has to be redetermined from the beginning.  In this
case, the number of {\it canonical pairs} may be $m < h$. 

A {\it singular} {\it scheme A} for $SU(3)\otimes U(1)$ corresponds, 
for example, to the following conditions on the {\it
invariants}:
\begin{eqnarray}
 \Jc_1' &\equiv& F_1(\Jc_1,\Jc_2,\Jc_3) = \Jc_1 - \frac{2}{3} \Jc^2_3 = 0 
\nome{A.7} \\
 \Jc_2' &\equiv& F_2(\Jc_1,\Jc_2,\Jc_3) = \Jc_2 - \frac{2}{9} \Jc^3_3 = 0
~~,  \nonumber
\end{eqnarray}
which imply $m=2$ (${\cal Q}_3$ and ${\cal P}_3$ missing), 
and $W_a=0$. Upon these conditions, the expression of
${\cal Q}_2$ in (\ref{A.4}) loses its meaning and has 
to be redefined. The new expression is
\Be
\fl
  {\Qc}_2'  = -\frac{\Ts 1}{\Ts 2} \arctan
     \frac{\Ts 2 S^2 N_{zz} - {\Jc}_3 (S^2-S_z^2)}{
           S \sqrt{\Ts - (S^2 - S_z^2)^2 + 4 \Ec N_{zz} (S^2 - S_z^2) 
                    - 4 S^2 N_{zz}^2 }
          } ~~.~~~~
\nome{A.8}
\Ee
\vspace{3mm}
The {\it scheme B}, i.e. the {\it canonical completion} of
{\it scheme  A}, is accomplished in general according to one of the
following possibilities:

{\bf Type 1}) no new variable is added to the {\it scheme A} and
$k$ independent function of the
{\it canonical  invariants} are put identically equal to
constants. $2h$ typical variables $P_i$, $Q_j$ are
identified to the variables of the {\it kernel}. 
The explicit expression  of the canonical
generators, in terms of the variables 
$P_i$,$Q_j$ of the  {\it scheme B}, is obtained by inverting
the functions of the {\it scheme A}.  Then, an arbitrary {\it fixed}
(with respect to the group parameters) 
canonical transformation $\Sc$ gives the expression  of
the generators in terms of $2h$ generic canonical variable
$p_i$,$q_j$ (in our cases, these are the already given 
expressions of the generators in terms of 
the {\it physical} variables. 
Since the phase space contains no submanifolds
invariant under the action  of $G_r$, these  realizations
are called {\it irreducible}, for any $k$-{\it uple} of
allowed constant values of the functions of the invariants.

{\bf Type 2}) a certain number $l$ ($l \leq k$) of 
canonical variables $Q_s$,
(called {\it  supplementary} variables) turn out to be
coupled, or are axiomatically coupled,  to $l$
of the invariants $\Jc_i$, building up $l$ new canonical
pairs,  while $h=k-l$ independent functions of the  {\it
invariants} are put identically equal to constants. Then the
generators, as functions of the {\it typical} variables
$P_i$,$Q_i$,$Q_s$ ($i,j=1,\ldots,h ~~s=1,\ldots,l$)
are obtained by inversion as before but do not
depend on  the {\it supplementary} variables $Q_s$. The
phase-space is $2h+2l$ dimensional and, containing the
submanifolds  $\Jc_t (p,q)=const.$ $(t=1,\cdots,k-l)$ 
as invariant submanifolds, corresponds to {\it non-irreducible}
realizations for any $(k-l)$-{\it uple} of
allowed constant values of the {\it invariant} functions
that have been constrained. 
   An arbitrary {\it fixed} (with respect to the group parameters)
canonical transformation $\Sc$ gives the expression of the 
generators in terms of $2h +2l$ generic canonical variables
$p_i$,$q_j$ (in the cases discussed in the present paper, 
this functional dependence is the starting point, and is 
defined by the {\it already given} expressions of the
generators in terms of the {\it physical} variables).

A particular case  of 
{\bf Type 2} is the so-called {\it complete}
realization, corresponding to $l=k$ and $(r+k)/2$ canonical
pairs. This realization is completely determined, locally,
by the group structure. In geometrical terms, the variables 
of the the {\it kernel} together with the $k$ {\it
supplementary} variables  define a local chart on the orbits
of the co-adjoint representation of  $G_r$.

{\bf Type 3}) an arbitrary number $v\leq n-h$ of pairs of 
canonical variables $Q_u$, $P_u$ ($u=1,\cdots,v$)
({\it  inessential} variables) turn out to exist, or are
axiomatically added, to the {\it scheme
A}. These variables  are not canonically coupled to {\it
invariants} nor share any functional  relation with
the variables of {\it scheme A}, so that they commute 
with all the variables 
considered up to now. Then, one proceeds, as for
{\bf Type 1} or {\bf Type 2}, 
by inverting the  functional dependence of the {\it
typical} variables on the generators and performing an
arbitrary {\it fixed} canonical  transformation which leads to the
generic form of the realization in terms of $2n$ canonical
variables $p_i$,$q_i$. Since the {\it inessential} variables
define (trivial) invariant submanifolds in phase-space,
these realizations are {\it non-irreducible}.  Note that
{\bf Types} {\bf 1}-and-{\bf 3}, or {\bf 2}-and-{\bf 3}, 
are mutually compatible.

\vspace{1mm}
The canonical realization of the $SU(3)\otimes U(1)$ group,
corresponding  to the three-dimensional isotropic  harmonic oscillator,
is characterized by a  {\it scheme B} which is a  {\it
canonical completion} of {\it type 2} ($m=2,~~l=1$)  of the {\it
singular scheme A} defined by conditions (\ref{A.7}).
Precisely, we have:

\vspace{0.5cm}
\Be 
\fl 
\Ba{l}
\mbox{
\begin{tabular}{||cc|c||} 
\hline \hline
    $P_1(\vec{r},\vec{\pi}) = S_z $                          & 
    $P_2(\vec{r},\vec{\pi}) = \sqrt{\vec S^2}$               &
    $P_3(\vec{r},\vec{\pi}) = {\Jc}_3 = \Ec/k $                \\
    $Q_1(\vec{r},\vec{\pi}) = \arctan\frac{\Ts S_y}{\Ts S_x}$& 
    $Q_2(\vec{r},\vec{\pi}) = \Qc_2' $                       &
    $Q_3(\vec{r},\vec{\pi}) = {\cal Q}_\Ec$               \\
\hline \hline
\end{tabular}
}
\\[8mm]
\mbox{
\begin{tabular}{||cc||} 
\hline \hline
    $ {\Jc}_1' \equiv {\cal F}_{1}=0$  & 
    $ {\Jc}_2' \equiv {\cal F}_{2}=0$ 
\\ 
    --                         &
    --                        \\
\hline
\hline
\end{tabular}
}
\Ea
\Ee 
\vspace{2mm}
where: 

\vspace{2mm} 
\noindent {\bf a)} 

\begin{eqnarray}
\fl
  {Q}_2(\vec{\pi},\vec{r}) = {\Qc}_2'
     &\equiv& -\frac{\Ts 1}{\Ts 2} \arctan
     \frac{\Ts 2 S^2 N_{zz} - {\Jc}_3 (S^2-S_z^2)}{
           S \sqrt{\Ts - (S^2 - S_z^2)^2 + 4 \Ec N_{zz} (S^2 - S_z^2) 
                    - 4 S^2 N_{zz}^2 }
          } 
\nome{A.9} \\
    &=& \frac{1}{2 s} \arctan \left[\frac{
                p_\theta \tan\theta}
               {s}\right] 
       +\frac{1}{4 s} \arctan \left[\frac{
               s^2-\Ec\mu r^2}
              {s \sqrt{2 \Ec \mu r^2 - k \mu r^4 - s^2}}\right] ~~,
\nonumber
\end{eqnarray}
\noindent {\bf b})~~ 
the {\it supplementary} variable $Q_3(\vec{\pi},\vec{r})$
has the expression
\begin{eqnarray}
  Q_3(\vec{\pi},\vec{r})\equiv{\cal Q}_\Ec 
      &=& \frac{1}{2}\sqrt{\frac{\mu}{k}}
          \arctan \left[ \sqrt{\frac{\mu}{k}}
               \frac{\Ec-k r^2}{\sqrt{\mu(2\Ec r^2- k r^4) - s^2}}
          \right]
~~,
\nome{A.10}
\end{eqnarray}
and 

\noindent {\bf c})~~ no {\it inessential} variables are present. 

\vspace{2mm}

This realization is clearly {\it non-irreducible}, due to
the fact that the {\it internal energy} $\Ec$ is 
an invariant which has not a {\it fixed} value.
The above expression of $\Qc_\Ec$ has been
determined  by solving the time independent Hamilton-Jacobi
equation, with $S_z$,$\sqrt{\vec S^2}$ and $\Ec$ assumed as
integration constants.


\section{Canonical Realization of the 
{\it Centrally Extended} Galilei Group ${\cal G}_M$}

The commutation relations of the {\it centrally extended} Galilei Lie
algebra are:
\Be
\left\{
\Ba{lcl}
  \{ \bar{J}_i, {J}_j \}     &=& \epsilon_{ijk} \bar{J}_k  \\
  \{ P_i      , {J}_j \}     &=& \epsilon_{ijk} P_k  \\
  \{ P_i      , {P}_j \}     &=& 0 \\
  \{ {K}_i, {J}_j \}       &=& \epsilon_{ijk} \bar{K}_k  \\
  \{ {K}_i, {K}_j \}       &=& 0  \\
  \{ \bar{H}  , {K}_i \}   &=& - P_i        \\
  \{ P_i      , {K}_j \}   &=& - M \delta_{ij}  ~~.
\Ea
\right.
\nome{B.1}
\Ee

\noindent The {\it scheme A} for the {\it centrally extended} 
Galilei group can be written:

\Be\mbox{
\begin{tabular}{||cc|cc||} 
\hline
\hline
    $ {\cal P}_i = P_i$                  &
    $ {\cal P}_4 = \Sigma_z$             &
    $ {\cal J}_1 = \sqrt{\vec \Sigma^2}$ &
    $ {\cal J}_2 = H - \frac{P^2}{2 M} $ \\ 
    $ {\cal Q}_i = \frac{K_i}{M} $       &
    $ {\cal Q}_4 = \arctan\frac{\Ts \Sigma_y}{\Ts \Sigma_x}$ & 
    &   \\
\hline
\hline
\end{tabular}
}\Ee
\vspace{0.2cm}
where $\vec{\Sigma} = \vec{J} - \frac{1}{M} \vec{K} \wedge \vec{P}$.

\vspace{2mm}
A {\it singular scheme A} for the {\it centrally extended} Galilei group  
obtains if ${\Jc}_1= 0$, which leads to $\vec{\Sigma} = 0$,
and therefore to the reduced {\it scheme} A:

\vspace{2mm}
\Be\mbox{
\begin{tabular}{||c|c||} 
\hline
\hline
    $ {\cal P}_i = P_i$                  &
    $ {\Jc}_2 = H - \frac{P^2}{2 M} $ 
\\ 
    $ {\cal Q}_i = \frac{K_i}{M} $   &
\\
\hline
\hline
\end{tabular}
}\Ee
\vspace{0.2cm}

\section{Canonical Realizations of the 
group~${\cal H}= {\cal G}_M  \wedge ~ ( SU(3)  \otimes U(1) ~)$}

The {\it scheme A} for the {\it direct product }
${\cal G}_M  \otimes SU(3)  \otimes U(1)$ is,
as for any other {\it direct product}, simply obtained by joining 
together the {\it schemes A} of the factor groups. Thus:

\vspace{0.5cm}
\Be 
\fl
\Ba{l}
\mbox{
\begin{tabular}{||ccc|cc||} 
\hline
\multicolumn{5}{||c||}{{\it kernel}} \\
\hline
\multicolumn{3}{||c|}{{\it SU(3)}} &
\multicolumn{2}{c||}{$\Gc_M$} \\
\hline \hline
    $ \Pc_1 = S_z $                         & 
    $ \Pc_2 = \sqrt{\vec S^2}$               &
    $ \Pc_3   $         &
    $ \Pc_{3+i} = P_i$             &
    $ \Pc_7 = \Sigma_z$        
\\ 
    $\Qc_1 = \arctan\frac{\Ts S_y}{\Ts S_x}$ & 
    $\Qc_2   $          &
    $\Qc_3   $          &
    $\Qc_{3+i} = \frac{K_i}{M} $       &
    $\Qc_7 = \arctan\frac{\Ts \tilde{\Sigma}_y}{\Ts \tilde{\Sigma}_x}$ 
\\
\hline
\hline
\end{tabular}
}
\\[12mm]
\mbox{
\begin{tabular}{||cc|c|cc||} 
\hline
\multicolumn{5}{||c||}{{\it invariants}} \\
\hline
\multicolumn{2}{||c|}{{\it SU(3)}} &
\multicolumn{1}{c}{{\it U(1)}} & 
\multicolumn{2}{|c||}{$\Gc_M$} \\
\hline
\hline
    ${\Jc}_1 $  & 
    ${\Jc}_2 $  &
    $\Jc_3   $  &
    ${\Jc}_4 = \sqrt{\vec {\Sigma}^2}$ &
    ${\Jc}_5 = H - \frac{P^2}{2 M}$ \\ 
    &
    & 
    &
    &
    \\
\hline
\hline
\end{tabular}
}
\Ea
\nome{quaCM}
\Ee
\vspace{0.2cm}

Consider now the {\it semi-direct product} of the {\it centrally extended}
Galilei group  and the $SU(3)\otimes U(1)$ group. 
The order of this group is $10+8+1=19$. 
Its Lie algebra is given by 
eq.(\ref{A.3}), (\ref{B.1}) {\it plus} the additional 
Poisson brackets which express 
the infinitesimal action  of ${\cal G}_M$ onto the
invariant subgroup  $SU(3)\otimes U(1)$; precisely:
\Be
\left\{
\Ba{rcl}
  \{ {J}_i ,{S}_a    \} &=& \epsilon_{iac}  {S}_c \\
  \{ {J}_i ,{N}_{ab} \} &=& \epsilon_{iac} {N}_{cb}
                               +\epsilon_{ibc} {N}_{ac}
~~.
\Ea
\right.
\nome{C.1}
\Ee

The {\it scheme A} for the group $\Hc$ can then be written:

\vspace{0.5cm}
\Be
\fl
\Ba{l}
\mbox{
\begin{tabular}{||ccc|cc||} 
\hline
\multicolumn{5}{||c||}{{\it kernel}} \\
\hline
\multicolumn{3}{||c|}{{\it SU(3)}} &
\multicolumn{2}{c||}{$\Cc$} \\
\hline \hline
    $ \Pc_1 = S_z $                & 
    $ \Pc_2 = \sqrt{\vec S^2}$     &
    $ \Pc_3   $                    &
    $ \Pc_{3+i} = P_i$             &
    $ \Pc_7 = \sigma_z$        
\\ 
    $\Qc_1 = \arctan\frac{\Ts S_y}{\Ts S_x}$ & 
    $\Qc_2   $          &
    $\Qc_3   $          &
    $\Qc_{3+i} = \frac{K_i}{M} $       &
    $\Qc_7 = \arctan\frac{\Ts {\sigma}_y}{\Ts {\sigma}_x}$ 
\\
\hline
\hline
\end{tabular}
}
\\[12mm]
\mbox{
\begin{tabular}{||cc|c|cc||} 
\hline
\multicolumn{5}{||c||}{{\it invariants}} \\
\hline
\multicolumn{2}{||c|}{{\it SU(3)}} &
\multicolumn{1}{c}{{\it U(1)}} & 
\multicolumn{2}{|c||}{$\Cc$} \\
\hline
\hline
    ${\Jc}_1 $  & 
    ${\Jc}_2 $  &
    $\Jc_3   $               &
    ${\Jc}_4 = \sqrt{\vec {\sigma}^2}$ &
    ${\Jc}_5 = H - \frac{P^2}{2 M}$ \\ 
    &
    &
    & 
    &
    \\
\hline
\hline
\end{tabular}
}
\Ea 
\nome{quaGAL}
\Ee
\vspace{0.2cm}
where $\Cc$ means the Galilei groups associated to
the {\it center-of-mass} coordinates, 
$\Qc_2$, $\Qc_3$, $\Pc_3$, $\Jc_1$, $\Jc_2$, $\Jc_3$ 
are given by eqs.(\ref{A.4}),(\ref{A.5}),({\ref{A.6}) and
$\vec{\sigma}$ is an angular momentum vector defined by: 
\Be
\vec{\sigma} \equiv \vec{\Sigma} - \vec{S} = \vec{J} 
- \frac{1}{M} \vec{K} \wedge \vec{P} - \vec{S} 
~~,
\Ee
so that
\Be
 \{ S_i , \sigma_j \} = 0 ~~~.
\Ee
Note that the only difference between the {\it schemes A}
described in (\ref{quaCM}) and (\ref{quaGAL}) 
is precisely due to the substitution
\Be
   \vec{\Sigma} \Rightarrow \vec{\sigma}
~~~,
\Ee
which accounts for the action of $\Gc_M$ onto $SU(3)$.
The variables of the {\it sub-scheme} $\Cc$ formally 
correspond to a {\it new centrally extended} Galilei group.

The canonical realization of $\Hc$ corresponding to the 
three dimensional isotropic harmonic oscillator with 
{\it center-of-mass} discussed in the present paper 
corresponds to a singular {\it scheme A}
in which four of the {\it invariant} are fixed 
in the following way:
\Bea
    \Jc_1^\prime &=& \Jc_1 - \frac{2}{3} \Jc_3^2 = 0 \nonumber \\
    \Jc_2^\prime &=& \Jc_2 - \frac{2}{9} \Jc_3^3 = 0 \nonumber \\
    \Jc_4        &=& 0 
~~~~~~\mbox{(which implies $\vec{\Sigma}\equiv\vec{S}$)}\\
    \Jc_5^\prime &=& \Jc_5 - \sqrt{\frac{k}{\mu}} \Jc_3 = 0 ~~. \nonumber
\Eea
These conditions lead to
 $m=5$ ($\Qc_3,~\Pc_3,~\Qc_7,~\Pc_7$ 
missing)
while  $\Qc_2$ has to be replaced by the expression given in eq. 
(\ref{A.10}). In conclusion, the {\it scheme B} of 
the realization has the form:

\vspace{0.5cm}
\Be 
\fl
\Ba{l}
\mbox{
\begin{tabular}{||cc|c||} 
\hline
\multicolumn{3}{||c||}{{\it kernel}} \\
\hline
\multicolumn{2}{||c|}{{\it SU(3)}} &
\multicolumn{1}{c||}{$\Cc$} \\
\hline \hline
    $ P_1(\vec{r},\vec{\pi}) = S_z $                         & 
    $ P_2(\vec{r},\vec{\pi}) = \sqrt{\vec S^2}$               &
    $ P_{3+i}(\vec{x},\vec{p}) = P_i$             
\\ 
    $ Q_1(\vec{r},\vec{\pi}) = \arctan\frac{\Ts S_y}{\Ts S_x}$ & 
    $ Q_2(\vec{r},\vec{\pi})   $          &
    $\Qc_{3+i}(\vec{x},\vec{p}) = \frac{K_i}{M} $ 
\\
\hline
\hline
\end{tabular}
}
\\[12mm]
\mbox{
\begin{tabular}{||cc|c|cc||} 
\hline
\multicolumn{5}{||c||}{{\it invariants}} \\
\hline
\multicolumn{2}{||c|}{{\it SU(3)}} &
\multicolumn{1}{c}{{\it U(1)}} & 
\multicolumn{2}{|c||}{$\Cc$} \\
\hline
\hline
    $ {\Jc}_1' \equiv {\cal F}_{1}=0$ & 
    $ {\Jc}_2' \equiv {\cal F}_{2}=0$ &
    $ \Jc_3 = \Ec/k $               &
    $ {\Jc}_4 = 0 $ &
    $ {\Jc}_5'= \Jc_5 - \sqrt{\frac{k}{\mu}} \Jc_3 = 0 $ \\ 
    --  &
    --  & 
    $ {Q}_\Ec(\vec{r},\vec{\pi}) $ &
    --  &
    --   \\
\hline
\hline
\end{tabular}
}
\Ea
\nome{genFIN}
\Ee
\vspace{0.2cm}

It is seen that (\ref{genFIN}) is a {\it scheme B} of
the {\it direct product} of  
$SU(3)\otimes U(1)$ times the {\it singular}
realization of $\Cc$ corresponding to the Galilei group 
transformations embodied by a scalar body ({\it center-of-mass}). 

Thus, the dynamical system we are describing has 11 
functionally independent
constants  of the motion, precisely: the {\it total momentum}
($\vec{P}$), a vector  ($\vec{R}=\frac{1}{M} \vec{K} - t
\vec{P}$) which expresses the inertial motion of the {\it
center-of-mass},  the {\it internal} angular momentum
($\vec{S}$) (which is equal to the {\it total} ``spin'' of the
system), the {\it internal} energy ($\Ec\equiv H -
\frac{\vec{P}^2}{2 M}$) and an {\it angular parameter} ($Q_2$) which fixes
the {\it principal axis} of the orbits. On the other hand, $\Qc_\Ec$
parametrizes the {\it evolution} on the orbits.


\def\vol#1{{\bf #1}}


\end{document}